\documentclass[%
reprint,
onecolumn,
showpacs,preprintnumbers,
 amsmath,amssymb,
aps,
prc,
]{revtex4}

\usepackage{graphicx}
\usepackage{dcolumn}
\usepackage{bm}
\usepackage{longtable}

\newcommand{\abinitio}{\textit{ab initio}}
\newcommand{\urel}{\ensuremath{U_\mathrm{rel}}}
\newcommand{\cdash}{\multicolumn{1}{c}{--}}

\begin{document}

\title{\textit{Ab initio} calculations of nuclear widths via an
  integral relation}

\author{Kenneth M. Nollett}
 \email{nollett@anl.gov}
\affiliation{
Physics Division, Argonne National Laboratory, Argonne, IL~~60439, USA
}

\date{\today}

\begin{abstract}
I describe the computation of energy widths of nuclear states using an
integral over the interaction region of \textit{ab initio} variational
Monte Carlo wave functions, and I present calculated widths for many
states.  I begin by presenting relations that connect certain
short-range integrals to widths.  I then present predicted widths for
$5\leq A \leq 9$ nuclei, and I compare them against measured widths.
They match the data more closely and with less ambiguity than
estimates based on spectroscopic factors.  I consider the consequences
of my results for identification of observed states in $^8$B, $^9$He,
and $^9$Li.  I also examine failures of the method and conclude that
they generally involve broad states and variational wave functions
that are not strongly peaked in the interaction region.  After
examining bound-state overlap functions computed from a similar
integral relation, I conclude that overlap calculations can diagnose
cases in which computed widths should not be trusted.

\end{abstract}


\maketitle

\section{Introduction}
\label{sec:introduction}

The last decade and a half have seen enormous progress in the
description of light nuclei as collections of interacting nucleons
with the same properties as in vacuum
\cite{PW01,pieper08,navratil09,roth04}.  After the formulation of
potentials that fit nucleon-nucleon scattering data with high
accuracy, \textit{ab initio} calculations of nuclear structure
demonstrated that the energy spectra of nuclei small enough for
converged calculations (mass number $A\lesssim 12$) can be understood
as arising from the vacuum nucleon-nucleon interaction \cite{WP02}.
The addition of a three-nucleon interaction with only two to four
parameters produces quantitative agreement with experiment for the
nuclear binding energies and spin-orbit splittings, including the
correct ground state ($J^\pi=3^+$ instead of $1^+$) for $^{10}$B
\cite{PPWC01}.

For nuclei with $A\leq 4$, methods using correlated hyperspherical
harmonic bases or Faddeev and similar formulations have solved
bound-state, scattering, and reaction problems quite successfully
\cite{deltuva08,kievsky11,viviani11,marcucci09}.  The \textit{ab
  initio} methods that have been developed for $A>4$ nuclei are suited
mainly to treatment of bound states, but there has been significant
progress on unbound states in recent years.  The Green's function
Monte Carlo (GFMC) method has been formulated as a particle-in-a-box
method and used to compute phase shifts in the $^5$He system with a
full realistic Hamiltonian \cite{NPWCH07}.  The no-core shell model
(NCSM) has been merged with the resonating group method (RGM) to
produce phase shifts and reaction cross sections in several systems
using effective two-body forces without explicit three-body terms
\cite{navratil10,navratil11-li6,navratil11-be7pg,navratil11-a5}.  The
coupled-cluster method has been combined with a Gamow shell model
basis to compute complex energies $E=E_R - i\Gamma/2$ of resonances
and thus their excitation energies $E_R$ and total widths $\Gamma$.
It has been applied to resonances of He isotopes and $A=17$ nuclei,
but it has not been used to produce cross sections
\cite{hagen07,hagen10}.  All of these approaches to scattering are
computationally challenging, and significant human and computer effort
is required for each individual system.

Even before the development of true scattering/reaction calculations
in $A>4$ nuclei, there were accurate energy calculations for many
unbound resonance states \cite{PW01,navratil09}, produced by
approximating them with bound (i.e., square-integrable) wave functions
and real energies.  This approach is successful for states
sufficiently narrow that important features of their structure can be
accommodated within the model space.  One way to understand these
calculations is as approximations to Gamow's decaying complex-energy
states \cite{gamow28} with energies near the real axis (i.e., small
widths).

Quantum Monte Carlo (QMC) calculations of square-integrable
``pseudo\-bound'' states begin with a variational Monte Carlo (VMC)
calculation, in which a complicated but closed-form wave function
containing large amounts of correlation is produced by minimizing the
energy expectation value with respect to many variational parameters
\cite{PW01}.  The computational effort in the VMC method lies in
computing energy expectation values by Monte Carlo integration over
all particle coordinates, and a square-integrable wave function is
necessary for both energy and normalization integrals to converge.
The variational ansatz incorporates square integrability through
particle correlations that decay exponentially at large separation,
and reasonable energies result because the long-range tails of the
true resonant wave function have small amplitudes relative to the
``interaction region'' where all nucleons interact.

The second step in application of the QMC method to nuclei is a GFMC
calculation.  This step also requires square-integrable wave
functions for evaluation of matrix elements.  GFMC takes the VMC wave
function as a starting point and evolves the Schr\"odinger equation
for imaginary values of the time $t=i\tau$ through a series of small
steps.  This evolution filters high-energy contamination out of the
wave function and for large $\tau$ leaves behind the lowest-energy
eigenstate contained in the VMC starting point.  For bound or narrow
states, this procedure ``converges'' at large $\tau$ to a unique
energy \cite{PPWC01,PVW02,PWC04}.  For broad states the GFMC
propagation drifts slowly toward the lowest available threshold; even
in very long GFMC propagations, the curve of energy \textit{versus}
$\tau$ for such states fails to flatten out at a ``converged'' energy.
Presumably this also occurs for the narrow states, but with a decline
too slow to have been noticed in calculations so far.

Because the calculated energies of narrow resonances in the QMC
methods are believed to be accurate, it is natural to ask whether
widths could be extracted from these calculations, avoiding the
considerable effort of explicit scattering/reaction calculations.
Attempts to correlate the width with the rate of decline of the
energy, $dE/d\tau$, in the late-time GFMC propagation failed, so the
rate may depend more on the starting wave function than on the
physical width~\cite{pieper-private}.

Even in explicit scattering calculations, some widths are effectively
too narrow to resolve by QMC methods.  Direct QMC calculations of
widths so far consist of energy calculations for a series of specified
boundary conditions \cite{NPWCH07}, so the smallest width that can be
computed must be larger than the energy resolution of the method.  As
an extreme example, the ground state of $^9$B has a total energy of
$-56.3$ MeV and a width of 0.5 keV, roughly 0.01\% of the total.
Since the typical precision claimed for GFMC energy calculations is
1\%, this width will not be amenable to calculation in this way.

Here I present a method to extract approximate widths from QMC wave
functions.  The basis of this approach is a relation between the
partial width in a specified breakup channel and an integral over the
``interaction region'' where all of the nucleons are close to each
other.  This relation has been used in the literature, though not (to
my knowledge) in explicit application to many-body wave functions.  It
has been used particularly in models of alpha
\cite{kadmenskii73,bugrov89} and proton
\cite{aberg97,aberg98,davids98,sonzogni99,davids00,esbensen01} decays
of heavy nuclei in order to avoid direct integration of the
Schr\"odinger equation to impractically large radii.  It is also
closely related to Green's function formulations of scattering theory.
There has been considerable interest in applying such formulations to
\textit{ab initio} calculations in the recent literature: they have
been used to extract asymptotic normalization coefficients (ANCs) from
many-body bound states computed in various approximations
\cite{akram90,lehman76,viviani05,nw11} and are coming into wider use
for scattering problems
\cite{barletta09,suzuki09,kievsky10,suzuki10,kievsky11,romero-redondo11}.
Preparation for future reaction calculations of that kind is a primary
motivation of the present work.

Although the integral relation can be used to compute the partial
width in any two-body decay channel if wave functions of the parent
and daughter nuclei are known, I confine my consideration in the
present study to nucleon emission.  I compute widths for all narrow
($\Gamma \lesssim 1$ MeV) one-nucleon decays to bound states in $5
\leq A \leq 9$ nuclei.  I also present results for several specific
broad states for which better theoretical information concerning
widths would be useful and for some with unbound final states that are
well-represented by pseudo\-bound wave functions.

The integral relation applied here can be used with either GFMC or VMC
wave functions, extending even to use of the same computer routines.
Application to GFMC requires more computation and additional
bookkeeping, so I have chosen in this initial study to use only VMC
wave functions.  This work represents the extension to unbound states
of the methods presented in Ref. \cite{nw11}.  The integral relation
used here may be used with other many-body methods, but it is
particularly well-suited to QMC wave functions because of its
formulation as a short-range integral and because of the transparent
treatment of fermion antisymmetry in the QMC methods.

The remainder of this paper is organized as follows: In
Sec.~\ref{sec:derivation} I motivate and define the integral method,
and I discuss its accuracy and its connection to overlap calculations.
In Sec.~\ref{sec:vmc} I describe the application of the integral
method to VMC wave functions.  In Sec.~\ref{sec:results} I present the
results of calculations for specific states, compare computed and
experimental widths, and consider how accurately widths may be
computed from spectroscopic factors.  Along the way I present new (and
likely improved) overlap calculations of VMC wave functions that may
prove useful for treatment of transfer reactions.  Finally in
Sec.~\ref{sec:conclusion} I summarize my results and briefly mention
future directions for this work.

\section{An integral relation for resonance widths}
\label{sec:derivation}

\subsection{The connection between widths and asymptotic normalizations}
\label{sec:plausibility-and-channel-setup}

Consider a many-body wave function $\Psi$ at energy $E$ above the
threshold for breakup into clusters 1 and 2 that have wave functions
$\psi_1$ and $\psi_2$ respectively and no internal angular momentum.
Assume also for the moment that only this breakup channel is open.
Since $\psi_1$ and $\psi_2$ are spinless, the orbital angular momentum
$l$ of their motion is equal to the total angular momentum $J$ of
$\Psi$.  Given appropriate boundary conditions, $\Psi$ at large
separations $r_{12}$ of the clusters is a linear combination of an
incoming wave proportional to $I_l(\eta,kr_{12})/r_{12}$ and an
outgoing wave proportional to $O_l(\eta,kr_{12})/r_{12}$, each
normalized to probability flux $\hbar k/\mu$ in its appropriate
direction at $r_{12}\rightarrow\infty$.  (For all special functions, I
follow the conventions of Refs.~\cite{abramowitz72} and
\cite{lanethomas}.)  These functions depend on the energy $E$ through
the wavenumber $k\equiv \sqrt{2\mu E}/\hbar$, where $\mu$ is the
reduced mass of the clusters, and through the Sommerfeld parameter
$\eta \equiv Z_1Z_2e^2\mu/\hbar k$, where $Z_1$ and $Z_2$ are the
charge numbers of the clusters.  $O_l$ and $I_l$ solve the radial
Coulomb-Schr\"odinger equation
\begin{equation}
  -\frac{d^2 u_l}{d \rho^2} + \left(\frac{l(l+1)}{\rho^2} + \frac{2\eta}{\rho}\right) u_l =  u_l
\end{equation}
with $\rho = kr_{12}$ and outward and inward flux, respectively, at
$r_{12}\rightarrow\infty$.  In terms of $I_l$ and $O_l$, $\Psi$
(assumed to have angular-momentum projection $m$, omitted from
subsequent expressions) is,
\begin{equation}
\label{eq:inandout}
  \Psi(r_{12}\rightarrow\infty) = C_l \psi_1\psi_2
  Y_{lm}(\mathbf{\hat{r}}_{12})\left[I_l(\eta,kr_{12}) - S_l(k) O_l(\eta,kr_{12})\right]/r_{12}.
\end{equation}
where $Y_{lm}(\mathbf{\hat{r}}_{12})$ is a spherical harmonic.  In
this single-channel case with both incoming and outgoing waves, the
normalization $C_l$ is arbitrary.  If $k$ and $\eta$ are real,
conservation of probability guarantees that the effect of scattering
is to multiply the outgoing wave $O_l$ by a complex phase factor
relative to the incoming wave $I_l$, so that the function $S_l(k)$ may
be written in terms of a real phase shift $\delta_l(k)$ as $S_l =
e^{2i\delta_l}$.  The function $S_l(k)$ is the single-channel case of
the $S$-matrix, which gives the amplitude and phase of outgoing waves
in terms of specified ingoing amplitudes, and it determines the cross
section uniquely.

Resonances in the scattering of clusters 1 and 2 occur at real
energies near poles of $S_l(k)$ and produce peaks in the scattering
cross section where $\tan\delta_l \simeq \Gamma/(E-E_R)$ around some
resonance with energy $E_R$ and width $\Gamma$.  $S$-matrix poles in
general occur at complex $k$ (and thus complex $E$).  It is evident
from Eq. (\ref{eq:inandout}) that at such a pole the wave function
takes the form
\begin{equation}
\label{eq:s-matrix-pole}
  \Psi(r_{12}\rightarrow\infty) = C_l^\prime\psi_1\psi_2 Y_l(\mathbf{\hat{r}}_{12})
  O_l(\eta,kr_{12})/r_{12},
\end{equation}
with only outgoing flux.  This is the same sort of decaying
complex-energy state originally formulated in Gamow's treatment of
alpha decay \cite{gamow28}.  The probability flux out of this state is
$\hbar k|C^\prime_l|^2/\mu$, so the normalization constant is no
longer arbitrary but carries information about the size of the
outgoing flux relative to the total wave function.  If $\Psi$ has been
normalized to unit total probability in some finite region, its
lifetime is inversely proportional to the outward probability flux.
By writing $\Psi$ in the form
\begin{equation}
  \Psi(r_{12}\rightarrow\infty) = C_l^{\prime} \psi_1\psi_2
  Y_l(\mathbf{\hat{r}}_{12}) \left[ \left[S_l(k)\right]^{-1} I_l -
    O_l\right]/r_{12},
\end{equation}
and defining $\Psi$ to have unit norm in some regularization scheme,
it can be shown that the residue of the $S$-matrix pole is
proportional both to the squared normalization constant
$|C_l^{\prime}|^2$ and to the imaginary part of the pole energy
\cite{dolinskii77,humblet61,okolowicz12}.  The relations among
$|C_l^{\prime}|^2$, the lifetime, and the pole location imply that the
width $\Gamma = \hbar^2 k |C_l^{\prime}|^2 /\mu$.  (Derivations that
deal rigorously with the normalization of $\Psi$ may be found in
Refs. \cite{dolinskii77,humblet61}.)  The width may thus be computed
from the normalization constant $C_l^{\prime}$.

Physically realized systems have real energies, so formulation of the
width in terms of complex-energy Gamow states is often inconvenient.
For the QMC methods, not only is the energy real, but the wave
functions are stationary waves, sums of incoming and outgoing waves
with zero total flux in each channel.  Again assuming a single open
channel of given angular momentum, a standing-wave solution is
asymptotically
\begin{equation}
\label{eq:standing-wave}
  \Psi(r_{12}\rightarrow\infty) =
  C_l\psi_1\psi_2Y_l(\mathbf{\hat{r}}_{12})\left[F_l(\eta,kr_{12}) +
    K_l(k) G_l(\eta,kr_{12})\right]/r_{12},
\end{equation}
for some different $C_l$ than before.  $F_l(\eta,kr_{12})$ is the
regular Coulomb function that satisfies $F_l(\eta,0) = 0$,
$G_l(\eta,\rho)$ satisfies the Wronskian relation
\[ \frac{d F_l(\eta,\rho)}{d\rho} G_l(\eta,\rho) 
- F_l(\eta,\rho)\frac{d G_l(\eta,\rho)}{d\rho}
= 1\ ,
\]
and the two are related to $O_l$ and $I_l$ by
\begin{eqnarray}
\label{eq:outgoing}
  O_l(\eta,\rho)  &= & G_l(\eta,\rho) + i F_l(\eta,\rho)\\
\label{eq:incoming}
  I_l(\eta,\rho)  &= & G_l(\eta,\rho) - i F_l(\eta,\rho).
\end{eqnarray}
The function $K_l(k)$ is the single-channel case of the $K$-matrix.
Eqs. (\ref{eq:inandout}), (\ref{eq:outgoing}), and (\ref{eq:incoming})
combine to give
\begin{equation}
\label{eq:K-from-S}
K_l(k) = i\frac{1-S_l(k)}{1+S_l(k)}
\end{equation}
 so that $K_l(k) = \tan\delta_l(k)$.  Eq.~(\ref{eq:standing-wave}) may
 then be written as
\begin{equation}
\label{eq:standing-wave-trig}
  \Psi(r_{12}\rightarrow\infty) = C_l^\prime
  \psi_1\psi_2Y_l(\mathbf{\hat{r}}_{12})\left[\cos\delta_l F_l(\eta,kr_{12}) 
    + \sin\delta_l
    G_l(\eta,kr_{12})\right]/r_{12}
\end{equation}
for some $C_l^\prime$ determined by an appropriate normalization of
$\Psi$.  At a resonance, $\delta_l = \pi/2$ and $d\delta_l/dE > 0$.
This corresponds to a pole of $K_l(k)$ on the positive real axis, so
that
\begin{eqnarray}
\label{eq:Gl-norm}
  \Psi(r_{12}\rightarrow\infty) 
 &=& C_l^\prime\psi_1\psi_2 Y_l(\mathbf{\hat{r}}_{12}) G_l(\eta,kr_{12})/r_{12}\\
 & = & \frac{1}{2}C_l^\prime \psi_1\psi_2 Y_l(\mathbf{\hat{r}}_{12})
    \left[ I_l(\eta,kr_{12}) + O_l(\eta,kr_{12})\right]/r_{12}.
\end{eqnarray}
The standing-wave solution at resonance is thus the sum of equal
inward and outward fluxes, each of magnitude $\hbar
k|C^\prime_l|^2/4\mu$.  This is the rate at which $\Psi$ decays
through the outgoing wave and is replenished by the ingoing wave.  In
analogy with the complex-energy case, it can be shown that the residue
of the $K$-matrix at the resonance pole is proportional to
$|C_l^\prime|^2$ for a wave function normalized to unity in a finite
region \cite{humblet70,humblet90,akram99}, up to corrections arising
from the choice of normalization volume.  These corrections are small
as long as the wave function has a much smaller amplitude in the
asymptotic region than in the interaction region
\cite{humblet61,humblet70,humblet90}, as is expected for a long-lived
resonance state.  Comparing Eqs.~(\ref{eq:s-matrix-pole}) and
(\ref{eq:Gl-norm}), it is unsurprising that the result of a rigorous
derivation is $\Gamma\approx\hbar^2 k |C_l^\prime|^2/ \mu$.

I have sketched the considerations that lead to the connection between
resonance width and wave function asymptotic normalization for the
case of a single open channel and spinless daughter nuclei.  These
considerations carry over directly to \abinitio\ wave functions, but
it is necessary to account for complications absent from the toy
model.

Most complications of the \abinitio\ case amount to additional
bookkeeping implied by multiple final-state channels, antisymmetry of
the wave function, and daughters with nonzero angular momenta.
Multiple types of decay products may be emitted, and their non-zero
internal angular momenta may couple in multiple ways.  The simple
right-hand sides of Eqs.~(\ref{eq:inandout}) and
(\ref{eq:standing-wave}) are replaced by sums over all breakup and $l$
channels of a given total angular momentum $J$ and parity $\pi$.  The
formalism to describe multi-channel wave functions in this case may be
found in many treatments of reaction theory
(e.g. Refs.~\cite{rodberg67,newton82,lanethomas,humblet90}).  The
coefficient $S_l(k)$ in Eq.~(\ref{eq:inandout}) is replaced by a
matrix $S_{ab}$ that gives the outgoing flux in channel $a$ produced
by unit incoming flux in channel $b$, and $K_l(k)$ in
Eq. (\ref{eq:standing-wave}) is similarly replaced by a matrix
$K_{ab}$ that gives the irregular-function amplitude in channel $a$
produced by unit regular-function amplitude in channel $b$.  If the
initial state in a reaction problem has amplitude $x_a$ in channel
$a$, then
\begin{equation}
\label{eq:kmatrix-multichannel}
\Psi(\mathrm{all}\ r_a\rightarrow\infty) = 
\sum_a\mathcal{A}_a\left[\psi_{a1}^{J_{a1}}\left[\psi_{a2}^{J_{a2}}Y_{l_a}(\mathbf{\hat{r}}_a)\right]_{j_a}\right]_{J}
\left\{x_{a}F_{l_a}(\eta_a,k_ar_a) + y_a G_{l_a}(\eta_a,k_ar_a)\right\}/r_a\ .
\end{equation}
In this expression, channel $a$ is characterized by daughter nuclei
with wave functions $\psi_{a1}^{J_{a1}}$ and $\psi_{a2}^{J_{a2}}$,
wave number $k_a$, Sommerfeld parameter $\eta_a$, and daughter
separation $r_a$.  The orbital angular momentum of this channel is
$l_a$, and the square brackets denote coupling of the daughter angular
momenta $J_{a2}$ and $l_a$ first to ``$jj$-coupled'' angular momentum
$j_a$ and then with $J_{a2}$ to form total angular momentum $J$.  (The
use of this coupling anticipates its later utility in defining
channels in QMC wave functions.)  The antisymmetrization operator
$\mathcal{A}_a$ carries out an antisymmetric sum over all partitions
of nucleons into daughters $a1$ and $a2$ with mass numbers $A_1$
and $A_2$ and multiplies by the normalization $\sqrt{A_1!A_2!/A!}$.
The index $a$ is taken to specify the daughter nuclei as well as the
channel quantum numbers $j_a$, $J_{a1}$, $J_{a2}$, $l_a$, and $\pi_a$.

Away from resonance, the irregular-function amplitudes are given by
$y_a=\sum_b K_{ab}x_b$.  At a pole of $K_{ab}$, all asymptotic
channels that are coupled to the resonance have wave functions
proportional to $G_{l_a}\!(\eta_a,k_ar_a)$. Then the corresponding
$x_a$ are irrelevant and may be set to zero.  If $\Psi$ at a pole is
normalized to unit probability within some finite volume that includes
the whole interaction region, the residue of $K_{aa}$ is proportional
to $|y_a|^2$ \cite{humblet70,humblet90}.  The partial width of the
resonance in channel $a$ is proportional to this residue, just as it
was in the single-channel case discussed above
\cite{newton82,akram99}.  It can be shown that the imaginary part of
the corresponding $S$-matrix pole energy is proportional to the sum of
these partial widths \cite{newton82}, so the pole residues give the
partial widths, whereas the $S$-matrix pole location gives only the
total width.

The channel radial functions defined by the summed terms in
Eq.~(\ref{eq:kmatrix-multichannel}) can be isolated from $\Psi$ by
projecting it onto the channel functions
\begin{equation}
  \label{eq:channel-definition}
  \Phi_a({\bm\xi}_{a1},{\bm\xi}_{a2},\mathbf{r}_a) 
  \equiv 
  \mathcal{A}_a\tilde{\Phi}_{a,p}(\mathbf{\bm\xi}_{a1},\mathbf{\bm\xi}_{a2},\mathbf{r}_a)\,
\end{equation}
where the channel function in a given partition $p$ of the nucleons
into clusters is
\begin{eqnarray}
\label{eq:channel-single-partition}
\tilde{\Phi}_{a,p}(\mathbf{\bm\xi}_{a1},\mathbf{\bm\xi}_{a2},\mathbf{r}_a) 
& \equiv &
\left[
  \psi_{a1}^{J_{a1}}(\mathbf{\bm \xi}_{a1}^p)
  \left[
  \psi_{a2}^{J_{a2}}(\mathbf{\bm \xi}_{a2}^p)
   Y_{l_a}(\mathbf{\hat{r}}_a)\right]_{j_a}\right]_{J}\,
\\
{\bm\xi}_{ai} & = & {\bm\xi}_{ai}^p \equiv \left\{\mathbf{r}_j
- \mathbf{r}_{ai}\right\},~j\in~ai\,.
\end{eqnarray}
The ${\bm\xi}_{ai}$ are the internal coordinates of
$\psi_{ai}^{J_{ai}}$, written in terms of nucleon coordinates in the
specified partition $p$ as differences ${\bm\xi}_i^p$ of the
coordinates $\mathbf{r}_j$ of nucleons within daughter $i$ and the
center of mass of daughter $i$.  The daughter centers of mass
$\mathbf{r}_{ai}$ are related to their separation $\mathbf{r}_a$ by
\begin{eqnarray}
\mathbf{r}_{a2}-\mathbf{r}_{a1} &=&  \mathbf{{r}}_a\\
  A_1 \mathbf{r}_{a1} + A_2 \mathbf{r}_{a2} & = &0\,.
\end{eqnarray}
Using bracket notation to indicate inner product in nucleon
spin-isospin space and integration over all nucleon coordinates
$\mathbf{R}=\{\mathbf{r}_j\}$, 
\begin{equation}
\label{eq:direct-overlap}
  \langle\Phi_a|\frac{\delta(r_a-r)}{r_a^2}|\Psi\rangle = 
\sqrt{\frac{A_1!A_2!}{A!}}\sum_p (-1)^p
    \int\left[
      \psi_{a1}^{J_{a1}}({\bm\xi}_{a1}^p)
      \left[
        \psi_{a2}^{J_{a2}}({\bm\xi}_{a2}^p)
        Y_{l_a}(\mathbf{\hat{r}}_a)\right]_{j_a}\right]_{J}^\dag
    \frac{\delta(r_a-r)}{r_a^2}\Psi d^{3A}R,  
\end{equation}
which for an exact solution of the Hamiltonian gives the overlap
function
\begin{eqnarray}
\label{eq:projection}
  R_a(r) &\equiv&  \langle\Phi_a|\delta(r_a-r)|\Psi\rangle/r^2\\
  & 
\stackrel{\textstyle =}{\scriptstyle r\rightarrow\infty} 
& 
\left\{x_aF_{l_a}\!(\eta_{a},k_ar) + y_aG_{l_a}\!(\eta_a,k_ar)\right\}/r\ .
\end{eqnarray}
It is this function that multiplies the daughter wave functions in
Eq.~(\ref{eq:kmatrix-multichannel}).  (Formally the integral extends
over only $3(A-1)$ coordinates, because center-of-mass motion is
irrelevant.  The wave functions used in nuclear QMC calculations are
translationally invariant, allowing the integral to be written as
extending over all coordinates; the effect of center-of-mass motion
cancels out when dividing computed quantities by the wave function
normalization.)

\subsection{Integral relations and asymptotic normalizations}
\label{sec:integral-method}

Eq.~(\ref{eq:Gl-norm}) suggests that the partial width of a
real-energy resonance state may be found by first computing $\Psi$ at
resonance and then projecting onto the desired channel and dividing by
$G_l/r$ to obtain the partial width
\begin{equation}
  \label{eq:simple-anc}
\Gamma_a 
= 
\frac{\hbar k_a}{\mu_a}\left|\frac{R_a(r) r}{G_l(\eta_a,k_ar)}\right|^2.
\end{equation}
This approach is useful when $\Psi$ can be computed accurately in the
large-$r$ region (as in Ref. \cite{okolowicz12}).  However, it often
happens that $\Psi$ is computed accurately in the interaction region,
but computation in the asymptotic region described by
Eq.~(\ref{eq:kmatrix-multichannel}) is difficult or inconvenient.

A more robust approach proceeds through a Green's function formalism.
One begins with the Schr\"odinger equation,
\begin{equation}
\label{eq:schroedinger}
(H - E)\Psi = 0,
\end{equation}
with $H$ the Hamiltonian operator and $E$ the energy.  To isolate a
particular channel $a$, as defined above, it is useful to choose a
partition $p$ of nucleons into the daughter nuclei in that channel and
write $H$ as a sum of parts internal to the daughters and parts that
depend on their relative motion.  Working in the center-of-mass frame,
\begin{equation}
\label{eq:h-breakdown}
H = T_{a,p} + U^{a,p}_\mathrm{rel} + H_1^{a,p} + H_2^{a,p} + V^a_C - V^a_C,
\end{equation}
where $T_{a,p}$ is the kinetic energy of relative motion of the
daughters.  The operator $H_i^{a,p}$ is the part of the Hamiltonian
(kinetic plus potential) involving only the coordinates
${\bm\xi}_{ai}^p$ within daughter $i$.  $U^{a,p}_\mathrm{rel}$
contains the remaining terms of the potential, consisting of
interactions between nucleons in daughter 1 and nucleons in daughter
2:
\begin{equation}
\label{eq:urel-definition}
U^{a,p}_\mathrm{rel} = \sum_{i\in a1;j\in a2} v_{ij}
 + \frac{1}{2}\sum_{i\in a1;j,k\in a2} V_{ijk}
 + \frac{1}{2}\sum_{i,j\in a1;k\in a2} V_{ijk}\,,
\end{equation}
This is {\em not} an effective interaction but rather the sum of all
terms of the many-body potential (with two-body terms $v_{ij}$ and
three-body terms $V_{ijk}$) linking the two daughters.  The
point-Coulomb interaction between daughters,
$V^a_C(r_a)=Z_{a1}Z_{a2}e^2/r_a$, is added and subtracted for reasons
that will become apparent later; it is zero if one of the daughters is
a neutron.

The energy $E$ may be similarly broken up into the sum of the daughter
internal energies $E_{ai}$ and the channel energy $E_a$:
\begin{equation}
  E = E_{a1} + E_{a2} + E_a.
\end{equation}
Then Eq. (\ref{eq:schroedinger}) becomes
\begin{equation}
  (T_{a,p} + U^{a,p}_\mathrm{rel} + V^a_C - V^a_C 
  + H^{a,p}_1 + H^{a,p}_2 - E_{a1} -E_{a2} - E_a)\Psi = 0.
\end{equation}
Rearranging terms and multiplying by the operator 
$[T_{a,p} + V^a_C - E_a]^{-1}$, we find
\begin{eqnarray}
\label{eq:broken-up}
\Psi & = & - \left[T_{a,p} + V^a_C - E_a\right]^{-1} 
\left(\urel^{a,p} - V^a_C\right)\Psi\\
 && - \left[T_{a,p} + V^a_C - E_a\right]^{-1} 
  \left(H_1^{a,p} + H_2^{a,p} - E_{a1} - E_{a2}\right)\Psi\ .
\nonumber
\end{eqnarray}

The next step is to project $\Psi$ onto channel $a$, as in
Eq. (\ref{eq:projection}).  Projection of Eq.~(\ref{eq:broken-up})
onto a channel function $\tilde{\Phi}_{a,p}$ of
Eq.~(\ref{eq:channel-single-partition}) gives
\begin{eqnarray}
  \label{eq:channel-projection}
  \left[
\tilde{\Phi}_{a,p}({\bm\xi}_1^p,{\bm\xi}_2^p,\mathbf{r}_a^\prime)\right]^\dag\Psi 
  &=& 
  -\left[\tilde{\Phi}_{a,p}(\mathbf{r}_a^\prime)\right]^\dag
     \left[T_{a,p} + V^a_C - E_a\right]^{-1}
     \left(\urel^{a,p} - V^a_C\right)\Psi\\
     && - \left[\tilde\Phi_{a,p}(\mathbf{r}_a^\prime)\right]^\dag 
     \left[T_{a,p} + V^a_C - E_a\right]^{-1} 
     \left(H^{a,p}_1 + H^{a,p}_2 - E_{a1} - E_{a2}\right)\Psi \nonumber
\end{eqnarray}
(abbreviating $\tilde\Phi_{a,p}$ on the right side by omitting the
cluster internal coordinates), but since
\[
(H^{a,p}_i - E_{ai})\psi_{ai}^{J_a}({\bm\xi}_i^p) = 0,
\]
the second line of Eq. (\ref{eq:channel-projection}) is zero, and
\begin{equation}
\label{eq:better-channel-projection}
  \left[\tilde\Phi_{a,p}(\mathbf{r}_a^\prime)\right]^\dag\Psi 
  = -\left[\tilde\Phi_{a,p}(\mathbf{r}_a^\prime)\right]^\dag
     \left[T_{a,p} + V^a_C - E_a\right]^{-1} 
     \left(\urel^{a,p} - V^a_C\right)\Psi\ .
\end{equation}

Now, the operator $\left[T_{a,p} + V^a_C - E_a\right]^{-1}$ takes
functions of nucleon coordinates $\mathbf{R}$ to functions of nucleon
coordinates $\mathbf{R}^\prime$ with different values of the
separation $\mathbf{r}_a$ but with the ${\bm\xi}_i^p$ untouched.  Its
application to a function $\phi_1(\mathbf{R})$ and projection onto a
second function $\phi_2(\mathbf{R}^\prime)$ may be written as an
integral over a Green's function that contains a product of Coulomb wave
functions:
\begin{equation}
\label{eq:greens-function}
\phi_2^\dag(\mathbf{R}^\prime)\left[T_{a,p} + V^a_C -
  E_a\right]^{-1}\phi_1(\mathbf{R}) =
\frac{2\mu}{\hbar^2k_a}\phi_2^\dag(\mathbf{R}^\prime) \int d^3
r_a\ \frac{F_{l_a}\!(\eta_a,k_ar_<)G_{l_a}\!(\eta_a,k_ar_>)}{r_< r_>}
Y_{l_a}(\mathbf{\hat{r}}_a^\prime)Y_{l_a}^\ast(\mathbf{\hat{r}}_a)
\phi_1(\mathbf{R}),
\end{equation}
following the usual notation that $r_<$ denotes the smaller of $r_a$
(a Jacobi coordinate specified for partition $p$ by $\mathbf{R}$) and
$r_a^\prime$ (specified by $p$ and $\mathbf{R}^\prime$), while $r_>$
denotes the larger of $r_a$ and $r_a^\prime$.  Rewriting
Eq.~(\ref{eq:direct-overlap}) in terms of
Eq.~(\ref{eq:better-channel-projection}) and applying
Eq.~(\ref{eq:greens-function}),  the result of integration over
$\mathbf{r}_a^\prime$ and antisymmetrization over partitions $p$ is
\begin{eqnarray}
\label{eq:full-integral-relation}
 \langle\Phi_a|\frac{\delta(r_a-r)}{r_a^2}|\Psi\rangle
 & =& -\frac{2\mu}{\hbar^2k_a}\left[\frac{G_{l_a}\!(\eta_a,k_ar)}{r} 
   \mathcal{A}_a\int_{r_a<r}
  \frac{F_{l_a}\!(\eta_a,k_ar_a)}{r_a}\left[\tilde\Phi_{a,p}(\mathbf{r}_a)\right]^\dag
       \left(\urel^{a,p} - V^a_C\right)\Psi\, d^{3A}R
  \right.\\\nonumber
&&   \left. + \frac{F_{l_a}\!(\eta_a,k_ar)}{r}
  \left\{B_{a,\infty} + \mathcal{A}_a\int_{r_a>r}
  \frac{G_{l_a}\!(\eta_a,k_ar_a)}{r_a} \left[\tilde\Phi_{a,p}(\mathbf{r}_a)\right]^\dag
       \left(\urel^{a,p} - V^a_C\right)\Psi\, d^{3A}R
  \right\}
  \right]\,.
\end{eqnarray}
The constant of integration $B_{a,\infty}$ can be determined from
$\Psi$ as in Ref. \cite{kievsky10,romero-redondo11}; as discussed
below, $B_{a,\infty}=0$ at resonance.  Finally, I denote the integrals
in Eq. (\ref{eq:full-integral-relation}) by
\begin{eqnarray}
\label{eq:f-norm-integral}
  B_a(r) & =&  \frac{2\mu}{\hbar^2k_a} \mathcal{A}_a\int_{r_a>r}
  \frac{G_{l_a}\!(\eta_a,k_ar_a)}{r_a} \left[\tilde\Phi_{a,p}(\mathbf{r}_a)\right]^\dag
  \left(\urel^{a,p} - V^a_C\right)\Psi\, d^{3A}R\\
\label{eq:g-norm-integral}
  C_a(r) &=& \frac{2\mu}{\hbar^2k_a} \mathcal{A}_a\int_{r_a<r}
  \frac{F_{l_a}\!(\eta_a,k_ar_a)}{r_a} \left[\tilde\Phi_{a,p}(\mathbf{r}_a)\right]^\dag
  \left(\urel^{a,p} - V^a_C\right)\Psi\, d^{3A}R\,,
\end{eqnarray}
so that
\begin{equation}
\label{eq:integral-schematic}
R_a(r) =
-\left\{\left[B_{a,\infty}+B_a(r)\right]F_l(\eta_a,k_ar) 
+ C_a(r)G_l(\eta_a,k_ar)\right\}/r.
\end{equation}
Many derivations of Eq. (\ref{eq:integral-schematic}) and equivalent
(through analytic continuation) bound-state expressions may be found
in the literature
(e.g. Refs.~\cite{rodberg67,pinkston65,timofeyuk98,kievsky10}).

Several observations may be made at this point.  The choice to add and
subtract the point-Coulomb interaction $V^a_C$ in
Eq.~(\ref{eq:h-breakdown}) has two important consequences.  First, it
guarantees that the overlap $R_a(r\rightarrow\infty)$ computed from
Eq.~(\ref{eq:integral-schematic}) is a linear combination of $F_l$ and
$G_l$, as required for solutions of the Schr\"odinger equation
governing $\Psi$.  Second, the nuclear interaction is short ranged, so
that at large separations, the interaction between daughter nuclei is
dominated by the monopole term of their Coulomb interaction.  Since
this is just equal to $V^a_C$, the difference $(\urel^{a,p}-V^a_C)$
goes rapidly to zero for $r$ beyond the range of the nuclear
interaction.  For the interactions and wave functions discussed below,
this typically happens at $r\simeq 7$ fm.  Thus, $B_a(r\geq
7\ \mathrm{fm})\sim 0$, and $C_a(r)$ takes on approximately its
asymptotic value for all $r \gtrsim 7$ fm.  These two properties make
Eq.~(\ref{eq:integral-schematic}) especially useful for computing
asymptotic properties of $\Psi$.  Indeed, similar integrals tend to
appear in scattering theory for exactly this reason.

Comparison of Eqs. (\ref{eq:integral-schematic}) and
(\ref{eq:standing-wave-trig}) shows that since
$B_a(r\rightarrow\infty)=0$,
\begin{equation}
\label{eq:kmatrix-integral}
\tan\delta_a = K_{aa} = \frac{C_a(r\rightarrow\infty)}{B_{a,\infty}},
\end{equation}
as discussed in many places in the literature
(e.g. \cite{newton82,kievsky10,barletta09}).  In fact,
Eq. (\ref{eq:kmatrix-integral}) may be improved upon significantly to
yield estimates of $K_{aa}$ that are second-order variational in
approximations to
$\Psi$~\cite{blatt49,newton82,kievsky10,romero-redondo11}.  (These
improvements require considerably more computation, so I do not pursue
them here.)  From the discussion in
Sec.~\ref{sec:plausibility-and-channel-setup}, at resonance there is
no contribution from the regular function so that $B_{a,\infty}=0$ and
$K_{aa}$ has a pole.  The residue of $K_{aa}$ at this pole (for $\Psi$
normalized to unity within some radius $r_\mathrm{norm}$ that contains
the interaction region) is
\begin{equation}
  \mathcal{R} = \frac{\hbar^2k_a}{2\mu_a}|C_a(r\rightarrow\infty)|^2\,,
\end{equation}
up to corrections of order
\begin{equation}
\label{eq:humblet-correction}
\epsilon=
\frac{\hbar^2}{2\mu_a}\left|r_\mathrm{norm} R_a(r_\mathrm{norm})\right|^2
\frac{d}{dE_a} 
\left[\frac{k_aG_{l_a}^\prime\!(\eta_a,k_ar_\mathrm{norm})}{
    G_{l_a}\!(\eta_a,k_ar_\mathrm{norm})}\right]\, ,
\end{equation}
discussed in Refs.~\cite{humblet70,humblet90}.  It is easy to show
from Eq.~(\ref{eq:K-from-S}) that the residue of the $S$-matrix pole
is twice the residue of the corresponding $K$-matrix pole, and thus
the partial width of the resonance in channel $a$ is given by
\begin{equation}
\label{eq:gamma-from-C}
  \Gamma_a =  \frac{\hbar^2k_a}{\mu_a}\left|C_a(r\rightarrow\infty)\right|^2,
\end{equation}
again up to corrections of order $\epsilon$.  Since the integral
defining $C_a$ (Eq.~(\ref{eq:g-norm-integral})) is short-ranged,
Eq.~(\ref{eq:gamma-from-C}) for the partial width may be formulated as
the square of a straightforward integral over all nucleon coordinates
without specifying a boundary.

This result for the width is easily generalized to include bound
states.  These occur at negative energy, $E_a = -|E_a|$, so that $k_a
= i|k_a|$ and $\eta_a=-i|\eta_a|$.  Since they must be
square-integrable, their overlap functions are asymptotically
\begin{equation}
\label{eq:bound-as-outgoing-wf}
R_a(r\rightarrow\infty) \propto
O_l(\eta_a,k_ar\rightarrow\infty)/r \propto e^{ik_ar}r^{-i\eta_a-1}.
\end{equation}
For purely imaginary $k_a$, $O_l$ is proportional to the Whittaker
function, $W_{-|\eta_a|,l+\frac{1}{2}}(2|k_a|r)$.  Working with
analytic continuations of the linearly-independent pair of functions
$F_l$ and $O_l$ instead of $F_l$ and $G_l$, an integral relation
analogous to Eq.~(\ref{eq:full-integral-relation}) may be derived.  It
yields a result for $R_a(r)$ that is guaranteed to have the correct
form (Eq.~(\ref{eq:bound-as-outgoing-wf})) at large $r$ even if it is
generated from an approximate $\Psi$ that does not solve the
Schr\"odinger equation exactly.  In fact, an early use of integral
relations of the form considered here was to produce overlaps with the
correct asymptotic behavior from asymptotically-incorrect Hartree-Fock
wave functions \cite{pinkston65,kawai67}.  They have been used more
recently to compute the asymptotic normalizations $\alpha_a$ defined
by
\begin{equation}
\label{eq:whittaker-asymptotic}
  R_a(r\rightarrow\infty) = \alpha_a W_{-|\eta_a|,l_a+\frac{1}{2}}(2|k_a|r)/r
\end{equation}
from shell-model and \abinitio\ wave functions
\cite{akram90,timofeyuk98,timofeyuk10,lehman76,friar88,viviani05,nw11}.

It is evident from the discussion above that a variational
approximation to the wave function $\Psi$ allows two calculations of
$R_a(r)$: one from the definition in Eq.~(\ref{eq:direct-overlap}) and
one from the integral relation of
Eq.~(\ref{eq:full-integral-relation}).  If $\Psi$, $\psi_{a1}^{J_a}$,
and $\psi_{a2}^{J_a}$ are all eigenstates of their respective
Hamiltonians, both methods yield the same result.  In a typical
application of many-body methods, $\Psi$ is quite accurate in the
interaction region but inaccurate where the daughter nuclei are widely
separated.  Eq.~(\ref{eq:direct-overlap}) yields $R_a(r)$ that is only
as accurate as $\Psi$ is at $r$ and is not guaranteed to have the
correct form at large $r$.  If $R_a$ is computed from
Eq.~(\ref{eq:full-integral-relation}), its value at any radius depends
only on the values of $\Psi$ within the interaction region (apart from
the integration constant $B_{a,\infty}$, which is nonzero only for
nonresonant open channels).  At large radius the correct asymptotic
shape is guaranteed.  It seems likely that since
Eq.~(\ref{eq:full-integral-relation}) incorporates the Hamiltonian
directly (through $\urel^{a,p}$) and not just through the approximate
$\Psi$, it is also more accurate at smaller $r$ than the
directly-computed $R_a(r)$ \cite{kawai67}.

\section{Application to variational wave functions}
\label{sec:vmc}

In the following calculations I use variational wave functions that
were computed by the VMC method.  They are approximate solutions for a
Hamiltonian consisting of the sum of the Argonne $v_{18}$ (AV18)
two-body \cite{WSS95} and Urbana IX (UIX) three-body \cite{PPCPW97}
interactions, which appear both in the computation of the wave
functions and in the operator $\urel^{a,p}$ used in the integral
relation.  In this section I describe enough of the structure of the
VMC $p$-shell wave functions to discuss their asymptotic properties,
I explain the implementation of the integral relation, and I finally
comment on the asymptotics of the wave functions.  The wave functions
are described in detail in Ref.~\cite{wiringa09}.

\subsection{Variational wave functions}

The VMC wave functions begin with the spin-isospin function
$\Phi_\alpha(0000)_{1234}$, in which the spins and isospins of the
first four particles are organized into a Slater determinant like
those of a filled, $\alpha$-particle-like $0s$ shell.  (The zeroes
denote the total angular momentum $J$, its projection $M$, the total
isospin $T$, and its projection $T_z$; this notation follows
conventions in previous QMC work and should not be confused with the
channel function $\Phi_a$).  The addition of further particles while
retaining antisymmetry requires spatial dependence in the wave
function, i.e., placing particles into the $p$-shell.  This is done
using orbitals $\phi_p^{LS[n]}(r_{\alpha i})$ defined in terms of the
separation $r_{\alpha i}$ of particle $i$ from the center of mass of
the ``alpha core.''  The orbitals are derived from Woods-Saxon
potential wells and coupled to form states of definite angular
momentum, parity, spatial symmetry, and isospin in an $LS$-coupled
basis, and they fall off exponentially at large $r_{\alpha i}$.  The
effects of pairwise interactions between nucleons (through the AV18
potential) are then accounted for using two-body scalar correlations
$f(r_{ij})$ that mainly account for particles' avoidance of the
repulsive core of the potential and are functions of the separation
$r_{ij}$ of particles $i$ and $j$.  It has been found advantageous to
allow different scalar correlations depending on whether particles $i$
and $j$ are both in the $s$-shell ($f^{ss}$), both in the $p$-shell
($f^{pp}$), or one in each ($f^{sp}$).  Finally, there are analogous
three-body scalar correlations ($f^{sss}$, $f^{spp}$) due to both two-
and three-body interactions.  All of these elements ($\alpha$ core,
$LS$ coupled orbitals, and scalar correlations) are antisymmetrized
among particle permutations to make up the Jastrow wave function:
\begin{eqnarray}
\label{eq:jastrow}
  \Psi_J &= \mathcal{A} \Bigg\{\! 
   & \left.\prod_{i<j<k \leq 4}f^{sss}_{ijk}\ 
     \prod_{n \leq 4}\ \prod_{5\leq m  < A}\ 
	\prod_{ m < q \leq A}\ f^{spp}_{nmq}
     \prod_{t<u \leq 4}f^{ss}(r_{tu})\right.  
     \nonumber\\
  &&    \times \prod_{i \leq 4}\ \prod_{5\leq j \leq A}\ f^{sp}(r_{ij}) 
	\prod_{5 \leq k < l\leq A}f^{pp}(r_{kl})
     \nonumber\\
  && \times \left.  \sum_{LS[n]} \Big( \beta_{LS[n]} 
     \Phi_A(LS[n]JMTT_{z})_{P} \Big) \right\} \ .
\end{eqnarray}
The index $P$ denotes a specific permutation of the particles into
$s$- and $p$-shells (subsequently antisymmetrized by the operator
$\mathcal{A}$) and the amplitudes $\beta_{LS[n]}$ specify the
admixtures of $p$-shell $LS$ states of Young diagram $[n]$.  In turn,
the configuration of good $L, S, [n], J, M, T,$ and $T_z$ is given by
\begin{eqnarray}
\label{eq:p-shell}
\Phi_A(LSJM[n]TT_{z})_P & = &
     \Phi_{\alpha}(0 0 0 0)_{1234}
     \prod_{5\leq i\leq A}
     \phi^{LS[n]}_{p}(r_{\alpha i})
     \nonumber \\
 &&  \times \left[ \left[\prod_{5\leq j\leq A}Y_{1m_j}(\hat{r}_{\alpha j})
	\right]_{LM_L} \otimes
      \left[\prod_{5\leq k \leq A}\chi_k (\frac{1}{2}m_k)\right]_{SM_S}
     \right]_{JM} \nonumber \\
 &&  \times \left[\prod_{5\leq l\leq A}\nu_{l}(\frac{1}{2}t_z)\right]_{TT_z} \ ,
\end{eqnarray}
where the spinors $\chi_i$ and $\nu_i$ specify the angular momentum
and isospin states of particle $i$.  Because the Hamiltonian includes
operator as well as scalar terms (i.e., it acts on particle spins and
isospins), there are operator correlations in addition to central
correlations.  These are accounted for by writing the variational
trial function as
\begin{equation}
\label{eq:trial-wf}
     \Psi_T =   {\cal S}\prod_{i<j}\left(1+U_{ij} 
                + \sum_{k\neq i,j} \tilde{U}^{TNI}_{ijk} \right) \Psi_J \ ,
\end{equation}
where $U_{ij}$ and $\tilde{U}_{ijk}^{TNI}$ are two- and three-body
operators, and $\mathcal{S}$ is a symmetrization operator, needed to
preserve the antisymmetry of $\Psi_J$ because the $U_{ij}$ and
$\tilde{U}_{ijk}^{TNI}$ do not commute amongst themselves.  The
operator correlations are of the form
\begin{equation}
     U_{ij} = \sum_{2\leq q\leq 6} \left[ \prod_{k\not=i,j}f^q_{ijk}({\bf r}_{ik}
              ,{\bf r}_{jk}) \right] u_q(r_{ij}) O^q_{ij} \ ,
\end{equation}
where $f_{ijk}^q$ is an operator-independent three-body correlation,
and the operators $\mathcal{O}_{ij}^q = {\bm\tau_i\cdot\bm\tau_j}$,
${\bm \sigma}_i\cdot{\bm \sigma}_j$, ${\bm \sigma}_i\cdot{\bm
  \sigma}_j {\bm \tau}_i\cdot {\bm \tau}_j$, $S_{ij}$, and $S_{ij}{\bm
  \tau}_i\cdot {\bm \tau}_j$ (where ${\bm\sigma}_i$ are nucleon spin
operators, ${\bm\tau}_i$, are isospin operators, and $S_{ij}$ is the
tensor operator), appear in the largest operator terms in the AV18
potential.  The $u_q(r_{ij})$, together with the scalar correlation
$f^{ss}(r_{ij})$, solve a set of coupled Euler-Lagrange equations with
coefficients that serve as variational parameters, discussed in
Ref. \cite{wiringa91}.  The central correlation $f^{ss}$ falls off
exponentially to reflect the strong binding of the $s$-shell
particles, as do the $u_q$ functions.

The $f^{sp}(r_{ij})$ and $f^{pp}(r_{ij})$ correlations are constructed
to approximate $f^{ss}(r_{ij})$ at small $r_{ij}$ but to approach
constant values at large $r_{ij}$.  This guarantees that where
particles approach each other closely the wave function is governed by
the nucleon-nucleon interaction, but that the correlation between
widely-separated particles is dominated by binding to a ``mean field''
accounted for in the $\phi_p^{LS[n]}$ orbitals.  Thus, the asymptotic
region of $\Psi_T$ is dominated by the $\phi_p^{LS[n]}$, which have
much longer tails than the $f^{ss}$.

All of the functions appearing so far in this section are specified as
functions of variational parameters, either explicitly or in the
differential equations solved to compute correlations.  The optimum
values of those parameters are found by searching the parameter space
to minimize the energy expectation value,
\begin{equation}
\label{eq:variational-energy}
  E \leq \frac{\left\langle\Psi_T|H|\Psi_T\right\rangle
  }{\left\langle\Psi_T|\Psi_T\right\rangle}\ ,
\end{equation}
with both numerator and denominator computed by Monte Carlo integrals
over the particle coordinates.  As a final step, the Hamiltonian is
diagonalized with respect to the $p$-shell configurations labeled by
$LS[n]$ to find the coefficients $\beta_{LS[n]}$.  Diagonalization
both improves the variational energy of the ground state of given
$J^\pi$ and $T$ and provides access to excited states.

Energies of unbound $p$-shell states can almost always be lowered by
making their wave functions more diffuse (closer to threshold).  This
is also often true of bound states, where variational energies (but
not GFMC energies) can lie artificially above breakup thresholds
because of shortcomings of the variational ansatz.  For both bound and
resonance states, this problem is addressed by constraining
variational parameter searches to keep charge radii close to
experimentally known ground-state charge radii.

The values of the variational parameters for all states used in the
present calculations were provided by R.~B.~Wiringa
\cite{wiringa-private}.  They are the results of calculations in bases
of good isospin in which individual nucleons typically cannot have
definite charges.  For convenience in defining the integral relations
for neutron or proton decays, I carry out the calculations below in a
basis of good nucleon charge (essentially an $m$-scheme in the
particle isospins) so that the emitted nucleon is definitely either a
neutron or a proton.  For given variational parameters, this is only a
change of representation and does not alter observables.  The
variational minimization has in all cases been carried out for the
state of lowest $T_z$ in each isomultiplet.  I compute widths of the
isobaric analogues of these states by using isospin rotations of the
minimized states rather than carrying out independent variational
minimizations.

\subsection{Overlap and integral-relation calculations}

Calculations of explicit overlap functions $R_a(r)$
(Eq.~(\ref{eq:projection})) and integral relations
(Eqs.~(\ref{eq:f-norm-integral}) and (\ref{eq:g-norm-integral}))
involve several common elements.  While integral relations are much
more time-consuming because they contain the operator $\urel^{a,p}$,
most of the computational tasks in organizing the calculation amount
to constructing the channel vector
$\tilde\Phi_{a,p}(\bm{\xi}^p_{a1},\bm{\xi}^p_{a2},\mathbf{r}_a)$ given
by Eq.~(\ref{eq:channel-definition}) and then contracting either
$\Psi_T$ or $(\urel^{a,p}-V^a_C)\Psi_T$ against it.  For nucleon
emission, I take $\psi_{a1}^{J_{a1}}$ to be the wave function of the
daughter nucleus and $\psi_{a2}^{J_{a2}}$ to be the spinor of the
emitted nucleon.

The integrals of Eqs.~(\ref{eq:direct-overlap}) and
(\ref{eq:g-norm-integral}) are computed by Monte Carlo integration
over the particle coordinates, using the same sampling algorithm that
has long been used to compute the energy expectation value,
Eq.~(\ref{eq:variational-energy}).  Sampling follows the Metropolis
algorithm, using the weight function
\begin{equation}
  W(\mathbf{R}) = \Psi_T^\dag(\mathbf{R})\Psi_T(\mathbf{R})\ ,
\end{equation}
where $\Psi_T(\mathbf{R})$ is the variational wave function for the
$A$-body parent nucleus at particle coordinates $\mathbf{R}$.  The
delta function in Eq.~(\ref{eq:direct-overlap}) is accounted for by
sampling all particle coordinates and sorting the samples into narrow
bins of specified channel radius; this builds up the entire function
$R_a(r)$ from a single Monte Carlo walk.  The normalization integral
needed to give $\psi_{a1}^{J_{a1}}$ unit norm is computed in the same
Monte Carlo walk by which the overlap or integral relation is
computed.

Only relative coordinates are used in the definitions of $\Psi_T$ and
$\tilde\Phi_{a,p}$, so the results of the calculations are all
explicitly translation invariant, and for operator
\begin{equation}
\label{eq:integral-operators}
\mathcal{M}=\left\{
\begin{array}{ll}
  \delta(r_a-r)/r_a^2 & \mbox{spectroscopic overlap}\\
  F_l(\eta_a,k_ar_a)(\urel^{a,p} -V^a_C)/r_a & \mbox{integral relation}
\end{array}\right.\ ,
\end{equation}
the quantity computed is
\begin{equation}
\label{eq:vmc-integrals}
I = \frac{\langle\Phi_a|\mathcal{M}|\Psi_T\rangle}{
  \langle\psi_{a1}^{J_{a1}}|\psi_{a1}^{J_{a1}}\rangle\langle\Psi_T|\Psi_T\rangle}\,.
\end{equation}
The routines used to compute the integral relations were written as
modified versions of existing spectroscopic-overlap routines
\cite{brida11,lapikas99,wuosmaa05-he7,wuosmaa05-li9}.  The
integral-relation routines were used previously to compute bound-state
ANCs \cite{nw11}, and only very minor modification (replacing regular
Whittaker functions with regular scattering functions) was necessary
for width calculations.

Finally, the operator $\urel^{a,p}$ must be considered.  It is just
the potential-energy operator of the AV18+UIX Hamiltonian, but with
the restriction that only terms involving the $p^\mathrm{th}$
(emitted) nucleon are considered so that
Eq.~(\ref{eq:urel-definition}) becomes:
\begin{equation}
\urel^{a,p} = \sum_{i\neq p} v_{ip} + \sum_{i<j,i\neq p,j\neq p} V_{ijp}\,.
\end{equation}
Its action on $\Psi_T$ is evaluated by calling the potential-energy
routines with instructions to omit all terms purely internal to
$\psi_{a1}^{J_{a1}}$ for the given permutation.  It is then
straightforward to contract $(\urel^{a,p} - V^a_C)\Psi_T$ with
$[\tilde\Phi_{a,p}(\bm\xi_{a1}^p,\bm\xi_{a2}^p,\mathbf{r}_a)]^\dag$
for a given configuration $\mathbf{R}$.

\begin{figure}
\includegraphics[width=3.5in,angle=0]{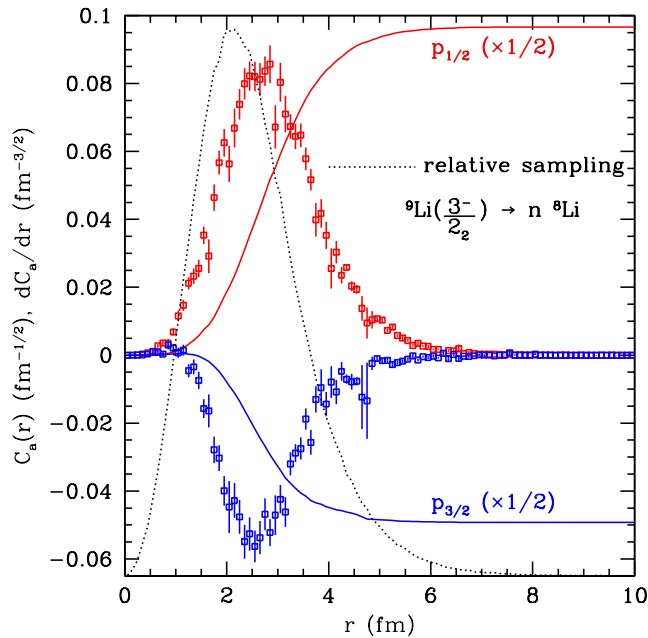}
  \caption{(Color online) The distribution of the integrand of
    Eq.~(\ref{eq:g-norm-integral}) in two $j_a$ channels is shown as a
    function of channel radius $r$ after integrating over all other
    coordinates.  The case shown is the $^9\mathrm{Li}(3/2^-_2)$
    overlap onto $^8\mathrm{Li}(\mathrm{g.s.})$ + neutron, but the
    shapes of the curves are similar in nearly all calculations from
    VMC wave functions.  Discrete points show the integrand $dC_a/dr$.
    The solid curve shows its cumulative integral, $C_a(r)$, which at
    large radius is equal to the wave function asymptotic
    normalization.  The integrand goes to zero at small radius because
    it contains the regular function $F_l(\eta_a,k_ar_a)$ and at
    $r_a\sim 7$ fm because $\urel^{a,p} - V^a_C \rightarrow 0$ there.
    The light dotted curve indicates the distribution of samples in
    the Monte Carlo integration (with zero at the bottom of the
    graph).}
  \label{fig:integrand}
\end{figure}

It is instructive to examine the integrand of
Eq.~(\ref{eq:g-norm-integral}) by inserting a delta function
$\delta(r_{a}-r)/r_a^2$ into the integral.  This is evaluated just
like the delta function in Eq.~(\ref{eq:direct-overlap}), by carrying
out the Monte Carlo walk for the full integral and binning the Monte
Carlo samples according to the channel radius instead of summing them
all together.  A typical result is shown in Fig. \ref{fig:integrand},
and nearly all cases that I have computed appear similar to this graph
apart from an overall scaling.  It is evident that the largest
contribution comes from $r_a \approx 2.5$ fm.  It is also evident that
the form of the VMC wave function beyond $\sim 7$ fm is irrelevant for
the width, because the factor $(\urel^{a,p}-V^a_C)$ is zero there.
This limits the integral to smaller radii and guarantees convergence
of the integral regardless of what lies in the tails of the
variational wave function.

The motivation presented above for the integral relations assumes that
$\Psi_T$, $\psi_{a1}^{J_{a1}}$, and $\psi_{a2}^{J_{a2}}$ are all
energy eigenstates.  To the extent that the VMC wave functions
approximate these states, the integral relations approximate the
overlaps and widths \textit{of the Hamiltonian for which they are
  approximate solutions}.  Two difficulties then present themselves in
applying the integral relations to VMC wave functions: 1) comparison
of the results to other calculations using the AV18+UIX Hamiltonian is
not possible because (except for $^5$He) no such calculations have
been done by other methods, and 2) comparison to experiment is
complicated because experimental resonance energies are not reproduced
exactly by the Hamiltonian.

The mismatch of resonance energies from experiment can be dealt with
straightforwardly.  The integral relations of
Eqs.~(\ref{eq:f-norm-integral}) and (\ref{eq:g-norm-integral}) require
an assumed channel energy $E_a$, from which $k_a$ and $\eta_a$ are
computed.  Formally $E_a$ should be the channel energy of the AV18+UIX
Hamiltonian (known from GFMC), but the mismatch between this and the
experimental $E_a$ is often $\sim 1$ MeV.  Since the potential energy
in the $p$-shell is much larger ($-100$ to $-400$ MeV), the
experimental $E_a$ is close enough to the AV18+UIX channel energy that
it can plausibly be used in the integrals.  (See Sec.~\ref{sec:psky}.)
This choice accounts for most of the well-known strong dependence of
the width on the resonance energy so that comparison with experiment
is possible.  It was found previously that using the experimental
separation energy in integral-relation calculations of bound-state
ANCs produces results in generally good agreement with experiment
\cite{nw11}.  Similarly, the results below indicate that using the
experimentally-measured resonance energy as $E_a$ allows prediction of
experimental widths.  I present the results of using both experimental
and GFMC values of $E_a$ in the integral relations.

More consistent calculations will require Hamiltonians that more
precisely reproduce thresholds and resonance energies.  Such
Hamiltonians exist in the form of the Illinois three-body potentials
\cite{PPWC01,pieper08-japan}, but they have not yet been incorporated
into the VMC code used here, and $\urel^{a,p}$ for these interactions
requires considerably more computation than for the UIX interaction.

\subsection{The asymptotic forms of VMC wave functions}

Consider configurations in which particle $A$ (before
antisymmetrization) is far from the first $A-1$ particles.  Because
$f^{pp}$ and $f^{sp}$ approach constants and $f^{ss}$ decays rapidly
at large $r_{ij}$, the shape of $\Psi_T$ in this part of the wave
function is dominated by the shapes of the single-particle functions
$\phi_p^{LS[n]}$.  This might be expected to give
\begin{equation}
\label{eq:overlap-from-orbitals}
R_a(r\rightarrow\infty) \approx \sum_{LS[n]} \gamma_{LS[n]}
\phi_n^{LS[n]}(\omega_{LS[n]}r)
\end{equation}
for some amplitudes $\gamma_{LS[n]}$ and Jacobian-like factors
$\omega_{LS[n]}$ that emerge from the correlations of
Eqs.~(\ref{eq:jastrow})-(\ref{eq:trial-wf}) when the overlap integral
is computed.  The factors $\omega_{LS[n]}$ account for the distinction
between the channel radius $r$ and the distance $r_{\alpha i}$ of a
$p$-shell particle from the center of mass of the $s$-shell core;
these only coincide when $A=5$ and otherwise differ by the mean
difference between nucleon-core and nucleon-daughter distances.

In general, the $\phi_p^{LS[n]}$ that emerge from the variational
procedure do not yield the correct long-range asymptotic shapes for
the overlaps $R_a$.  This is most readily seen for open channels,
where solutions that should oscillate at large channel radius instead
fall off with an assumed exponential dependence.  In closed channels,
the condition of square integrability gives zero for the analogue of
$B_{a,\infty}$ in Eqs.~(\ref{eq:full-integral-relation}) and
(\ref{eq:integral-schematic}), so that
Eq.~(\ref{eq:whittaker-asymptotic}) holds for the true wave function.
Because no single $LS[n]$ term typically dominates a given $R_a(r)$,
it is in general difficult to construct $\phi_p^{LS[n]}$ to satisfy
Eq.~(\ref{eq:whittaker-asymptotic}) for all possible channels at once.
This difficulty is compounded by the problem that the energy
expectation value driving the variational minimization receives very
little contribution from the wave function tails, so the variational
principle does not constrain the low-probability tails of the wave
function very strongly.  The application of the integral approach to
bound state ANCs in Ref.~\cite{nw11} avoided these difficulties and
effectively matched the correct asymptotic form onto the
better-computed interior of the wave function.

\begin{figure}
\includegraphics[width=3.5in,angle=0]{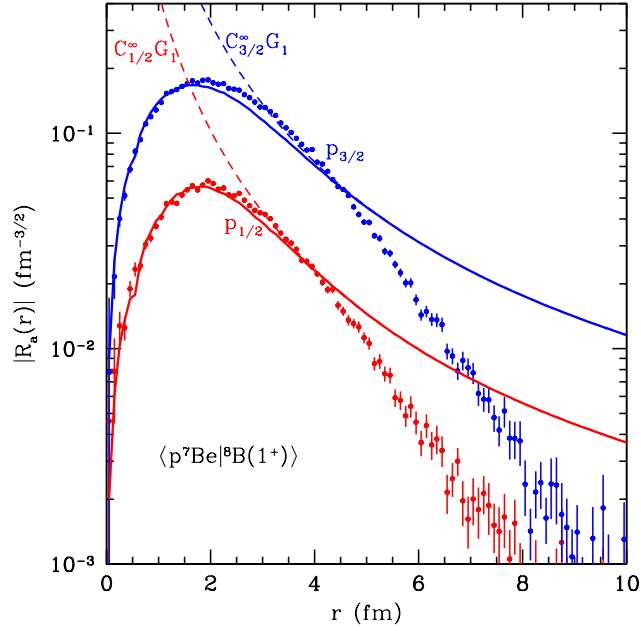}
  \caption{(Color online) Overlaps $R_a(r)$ of the unbound $J^\pi=1^+$
    state of $^8$B onto $^7\mathrm{Be}\ +$ proton, computed directly
    from Eq.~(\ref{eq:direct-overlap}) (discrete points with Monte
    Carlo error bars) and from the integral relation of
    Eq.~(\ref{eq:full-integral-relation}) (solid curves).  At large
    radius, the directly-computed overlap falls off exponentially
    because the $\phi_p^{LS[n]}$ in the VMC wave function fall off
    exponentially.  However, the integral-relation overlap (with
    $B_{a,\infty}=0$) gives the exact proportionality to $G_l$
    expected at resonance.  The absolute normalization
    $C^\infty_{j_a}\equiv C_a(r\rightarrow\infty)$ multiplying $G_l$
    gives the partial width of $^8\mathrm{B}(1^+)$ in each
    $^7\mathrm{Be}+p$ channel; the normalized asymptotic is shown as a
    dashed curve for each channel and manifestly merges with the full
    overlap at $\gtrsim 4$ fm.}
  \label{fig:fixing-asymptotics}
\end{figure}

Pseudo\-bound VMC wave functions approximate resonance states with
square-integrable wave functions, in which the $f^{ss}$ and
$\phi_p^{LS[n]}$ functions cut off the wave function tails
exponentially.  This exponential cutoff can be understood at
``medium'' range (4--8 fm) as forcing a resonance form on the wave
function and beyond this range as providing a regularization to
normalize the unbound state despite its formally nonzero amplitude at
large radius.  Such a regularization is important for quantities like
electromagnetic transition strengths (e.g.~\cite{pervin07,marcucci08})
and for the approximate relation between the asymptotic normalization
and the partial width given in Eq.~(\ref{eq:gamma-from-C})
\cite{humblet90}.  As long as the resonance wave function is computed
reasonably accurately within the region where the integral of
Eq.~(\ref{eq:g-norm-integral}) is nonzero, and it is normalized to
unity over the region where $\Psi_T$ is larger in amplitude than in
the asymptotic region, the integral approximates the asymptotic
normalization of $G_l$ and thus the partial width.  The cutoff of
$\Psi_T$ at large nucleon separation is illustrated with overlap
functions in Fig.~\ref{fig:fixing-asymptotics}.

\section{Results}
\label{sec:results}

I now apply the integral relation of Eqs.~(\ref{eq:gamma-from-C}) and
(\ref{eq:g-norm-integral}) to compute widths of several unbound energy
levels in nuclei of mass numbers $5\leq A \leq 9$, from VMC wave
functions.  I mostly choose energy levels that empirically have small
width (under 1 MeV) and are dominated by nucleon emission.  The
integral relation is valid for decay channels in which all three wave
functions (one parent and two daughters) are known, but I limit this
first examination to nucleon-emission channels, where one ``daughter
nucleus'' is a neutron or proton.  I concentrate on two-body final
states, but I also model some three-body decays as sequential
processes, e.g. the $^9$B ground state decaying to a proton and the
unbound ground state of $^8$Be.  The subsequent decay of $^8$Be to two
alpha particles can be neglected because of its small width.  (In
principle the widths of unbound daughter states should be integrated
over, but I find that in all cases considered this correction is much
smaller than either the experimental errors or the widths of omitted
decay channels.)

The results of the width calculations are shown in Table
\ref{tab:big-width-table}.  Each calculation was carried out twice:
once assuming the channel energy $E_a$ from the AV18+UIX Hamiltonian
(known from GFMC calculations) and once using the experimentally-known
resonance energy.  (For the second $J^\pi=2^+$ states in $^8$B and
$^8$Li and the $J^\pi=7/2^-$ states in $^9$Be and $^9$B, AV18+UIX
energies have never been computed with GFMC).  Where the experimental
channel energy is unknown or uncertain, I have used instead the GFMC
energy for the AV18+IL7 Hamiltonian, which gives a better overall fit
to experimental energies than AV18+UIX.

I have taken the experimental energies and widths in Table
\ref{tab:big-width-table} mainly from data compilations
\cite{eval-A=5-7,eval-A=8-10}.  Nearly all of the widths in the
compilations are ``observed'' widths, which coincide with the full
width at half maximum (FWHM) of cross section peaks and are
proportional to the sum of the $K$-matrix pole residues.  The
integral-relation widths should correspond to these quantities.  In
two cases ($^7\mathrm{Li}(5/2^-_2)$ and $^9\mathrm{Be}(1/2^-)$),
examination of the source literature indicated that the available
experimental widths are ``formal'' widths of the $R$-matrix formalism.
I have converted the experimental widths listed for those states to
``observed'' widths by the usual relation
\begin{equation}
\label{eq:observed-width}
  \Gamma_\mathrm{obs} = \frac{2 \sum_a \gamma_a^2 P_{l_a}(k_a b)}{1+\sum_a\gamma_a^2S_{l_a}^\prime(E_a)},
\end{equation}
where $b$ (taken to be 4 fm) is a matching radius, $S_l(E)$ is the
shift function of $R$-matrix theory \cite{lanethomas}, prime denotes
its derivative, and $\gamma_a^2$ is defined by the formal width
$\Gamma_a\equiv 2 \gamma_a^2 P_{l_a}\!(k_a b)$ and the penetrability
function $P_{l_a}(k_a r)$.  (See the section ``Definitions of resonance
parameters'' in Ref.~\cite{eval-A=5-7}.)  The sources of the
experimental widths are indicated in Table~\ref{tab:big-width-table}.

The widths presented in the table are sums over all $p$- and $f$-wave
decay channels.  The predicted $f$-wave contributions are less than
1\% of the total in all cases except the decay of
$^9\mathrm{Li}(7/2^-)$ to $^8\mathrm{Li}(2^+)$, where it is computed
to be 23\% of the total.  The $p$-wave decays are in most cases sums
of $p_{1/2}$ and $p_{3/2}$ contributions.  The decompositions into
$p_{1/2}$ and $p_{3/2}$ (or into a channel-spin coupling scheme) are
available on request to the author.

The table only includes cases for which established experimental or
GFMC energies are available.  I have also computed widths of several
states of $^9$He with varying assumptions about resonance energies as
described below, but I have not included these numbers in Table
\ref{tab:big-width-table}.

\begin{table}
  \caption{The results of integral-relation calculations of widths.
    Results are shown from calculations in which the channel energy
    was assumed equal to its experimental value (``From Expt energy'')
    and to the AV18+UIX channel energy known from GFMC (``From
    AV18+UIX energy'').  Error estimates are quadrature sums of Monte
    Carlo sampling errors and uncertainties from the input energies.
    Where no experimental energy is available, the results in the
    ``Experiment'' columns were computed using the GFMC energy with
    the AV18+IL7 Hamiltonian, and they are indicated by square
    brackets.  The column ``Matches $\pi/2$?''  indicates whether the
    overlap function appears consistent with a resonance state, as
    discussed in Sec.~\ref{sec:psky}.  See Sec.~\ref{sec:psky} and
    Eq.~(\ref{eq:zeta}) for the definition of $\zeta$.  Energies are
    relative to the decay threshold in the center-of-mass frame.
    Experimental energies are taken from data compilations
    \cite{eval-A=5-7,eval-A=8-10} unless otherwise noted.}
  \label{tab:big-width-table}
  \begin{tabular}{lllllllcl}
\hline
    State  & Daughter
& \multicolumn{2}{c}{Experiment} &   \multicolumn{1}{c}{From Expt energy} & \multicolumn{2}{c}{From AV18+UIX energy} & Matches & \multicolumn{1}{c}{$\zeta$} \\
&& 
$E$ (MeV) & $\Gamma$ (MeV)  &   $\Gamma_{VMC}$ (MeV) & $E_\mathrm{UIX}$ (MeV) &  $\Gamma_{VMC}$ (MeV) &  $\pi/2$? &     \\\hline
$^5\mathrm{He}(3/2^-)$ & $^4\mathrm{He}(0^+)$
 & 0.798 &  0.648 \cite{eval-A=5-7}   & 0.307(5)      &       1.39            &    0.684(11)   &  no & 0.460\\
$^5\mathrm{He}(1/2                 ^-)$ & $^4\mathrm{He}(0^+)$
 & 2.07  &  5.57  \cite{eval-A=5-7}  & 0.582(13)     &       2.4             &    0.711(15)   &  no  & 0.429 \\
$^7\mathrm{He}(3/2^-)$ & $^6\mathrm{He}(0^+)$
 & 0.445(3) &  0.122(13)\footnotemark[1] & 0.114(12)                       &       1.68(13) &    0.77(10)    & yes  & 0.092\\\hline
$^7\mathrm{He}(1/2^-)$ & $^6\mathrm{He}(0^+)$
 & 3.05(10)\footnotemark[2] &   \cdash    & 1.98(12)            &       2.83(13)            &    1.80(12)     &  no & 0.21\\
$^7\mathrm{He}(1/2^-)$ & $^6\mathrm{He}(2^+)$
 & 1.25(10)\footnotemark[2] &   \cdash    & 0.42(6)            &       0.89(13)  &    0.26(5)     & yes & 0.067\\
$^7\mathrm{He}(1/2^-)$ & sum
 &           & 2.0(1.0)\footnotemark[3] & 2.40(12)\footnotemark[4]  & 2.83(13)        &    2.22(11)\footnotemark[4]    &\\\hline
$^7\mathrm{He}(5/2^-)$ & $^6\mathrm{He}(2^+)$
 & 1.55(3)\footnotemark[2] & 1.99(17) \cite{eval-A=5-7} & 1.29(12)\footnotemark[4]  & 1.87(13)          &    1.7(2)\footnotemark[4]    &  no & 0.165\\
$^7\mathrm{Li}(5/2^-_2)$ & $^6\mathrm{Li}(1^+)$
 & 0.204 & 0.065(3) \cite{eval-A=5-7}   & 0.0483(17)\footnotemark[4]& 1.57(17)            &    0.92(13)\footnotemark[4]     & yes & 0.055\\
$^7\mathrm{Be}(5/2^-_2)$ & $^6\mathrm{Li}(1^+)$
 & 1.60(6) &  0.19(5) \cite{skill95} & 0.43(4)\footnotemark[4]&  2.65(17)                &    1.11(14)\footnotemark[4]     & yes & 0.055\\\hline
$^8\mathrm{Li}(3^+)$ & $^7\mathrm{Li}(3/2^-)$
 & 0.223(3) & 0.032(3) \cite{mughabghab73} & 0.0344(18)               &  2.10(18)             &    0.88(11)        & yes   & 0.007\\
$^8\mathrm{Li}(0^+)$ & $^7\mathrm{Li}(3/2^-)$
 & \![0.97(13)]& \cdash  & \![0.37(7)]                 &  0.67(17)            &    0.24(8)   & no & 0.005\\
$^8\mathrm{Li}(0^+)$ & $^7\mathrm{Li}(1/2^-)$
 & \![0.847(14)]& \cdash  & \![0.81(2)]               &  0.78(17)            &    0.7(2)     & no & 0.004\\\hline
$^8\mathrm{Li}(2_2^+)$ & $^7\mathrm{Li}(3/2^-)$
 & [2.18(16)]   & \cdash   &    [1.00(11)]                    &   \cdash    &    \cdash   & yes & 0.004\\
$^8\mathrm{Li}(2_2^+)$ & $^7\mathrm{Li}(1/2^-)$
 & [2.06(19)]   & \cdash   &   [0.105(14)]                   &   \cdash    &    \cdash   & yes & 0.010\\
$^8\mathrm{Be}(1^+)$ $T=1$\footnotemark[5] & $^7\mathrm{Li}(3/2^-)$
 & 0.385(1)  & \cdash  & 0.0089(3)                   &  1.26(19)           &    0.17(5)     & yes & 0.003\\
$^8\mathrm{Be}(1^+)$ $T=0$\footnotemark[5] & $^7\mathrm{Li}(3/2^-)$
 & 0.895(5)   & \cdash  & 0.152(4)                    &  0.51(21)          &    $0.04^{+0.05}_{-0.03}$   & yes & 0.003\\
$^8\mathrm{Be}(1^+)$ sum\footnotemark[5] & $^7\mathrm{Li}(3/2^-)$
 &         & 0.149(6) \cite{eval-A=8-10}  & 0.161(4) &               &    0.21(5)     & yes\\\hline
$^8\mathrm{Be}(3^+)$ $T=1$\footnotemark[5] & $^7\mathrm{Li}(3/2^-)$
 & 1.81(3)    & \cdash  & 0.166(9)                    &  3.68(18)           &     0.60(7)     & yes & 0.007\\
$^8\mathrm{Be}(3^+)$ $T=0$\footnotemark[5] & $^7\mathrm{Li}(3/2^-)$
 & 1.98(1)    & \cdash  & 0.314(14)                   &  2.3(2)            &    0.43(7)     & yes & 0.003\\
$^8\mathrm{Be}(3^+)$ $T=1$\footnotemark[5] & $^7\mathrm{Be}(3/2^-)$
 & 0.17(3)    & \cdash  & 0.012(3)                  &  2.14(18)            &    0.45(5)     & yes & 0.007\\
$^8\mathrm{Be}(3^+)$ $T=0$\footnotemark[5] & $^7\mathrm{Be}(3/2^-)$
 & 0.335(10)   & \cdash  & 0.050(3)                   &  0.8(2)            &    0.18(6)    & yes & 0.004\\
$^8\mathrm{Be}(3^+)$ sum\footnotemark[5] & sum
 &          & 0.50(3) \cite{eval-A=8-10} & 0.542(17) &                &    1.66(13)     & yes\\\hline
$^8\mathrm{B}(1^+)$ & $^7\mathrm{Be}(3/2^-)$
 & 0.632(3) & \cdash   & 0.0382(15)              &  1.3(3)            &    0.26(13)     & yes & 0.001\\
$^8\mathrm{B}(1^+)$ & $^7\mathrm{Be}(1/2^-)$
 & 0.203(3) & \cdash   & 0.00105(8)              &  1.4(2)            &     0.5(2)     & yes & 0.003\\
$^8\mathrm{B}(1^+)$ & sum
 &       & 0.0357(6) \cite{eval-A=8-10} & 0.0394(15)  &           &    0.8(2)     & yes &       \\\hline
$^8\mathrm{B}(3^+)$ & $^7\mathrm{Be}(3/2^-)$
 & 2.18(2)  & 0.39(4)\footnotemark[6] & 0.38(2)\footnotemark[4]   &  3.7(2)        &    1.08(11)\footnotemark[4] & yes   & 0.007\\
$^8\mathrm{B}(0^+)$ & $^7\mathrm{Be}(3/2^-)$
 & \![2.55(13)]& \cdash  & \![0.65(7)]                 &  2.17(17)                 &    0.47(8)     & no & 0.005\\
$^8\mathrm{B}(0^+)$ & $^7\mathrm{Be}(1/2^-)$
 & \![2.44(14)]& \cdash  & \![1.46(18)]                &  2.30(13)            &    1.3(2)     & no & 0.004\\
$^8\mathrm{B}(2_2^+)$ & $^7\mathrm{Be}(3/2^-)$
 & 2.41(2) \cite{mitchell10}     &  0.12(4) \cite{mitchell10} &    0.51(2)              &   \cdash    &    \cdash   & yes & 0.004\\
$^8\mathrm{B}(2_2^+)$ & $^7\mathrm{Be}(1/2^-)$
 & 1.98(2) \cite{mitchell10}    &  0.24(11) \cite{mitchell10} &   0.039(2)               &   \cdash    &    \cdash   & yes & 0.010\\
$^9\mathrm{Li}(5/2^-)$\footnotemark[7] & $^8\mathrm{Li}(2^+)$
 & 0.232(15)    & 0.10(3) \cite{wuosmaa05-li9}  & 0.145(14)                   &  0.98(44) &  1.2(8)    & yes & 0.003\\\hline
$^9\mathrm{Li}(7/2^-)$\footnotemark[7] & $^8\mathrm{Li}(2^+)$
 & 2.366(15)    & \cdash & 0.0012(5)                   &  3.6(3)           &    0.0031(13) & no & 0.045 \\
$^9\mathrm{Li}(7/2^-)$\footnotemark[7] & $^8\mathrm{Li}(3^+)$
 & 0.111(15)  & \cdash & 0.043(8)                   &  0.23(35)            &    $< 0.50$   & yes & 0.006\\
$^9\mathrm{Li}(7/2^-)$\footnotemark[7] & sum
 &          & 0.04(2) \cite{wuosmaa05-li9} & 0.044(8)\footnotemark[4]      &              &      $< 0.50$\footnotemark[4] &  \\\hline
$^9\mathrm{Li}(3/2^-_2)$\footnotemark[7] & $^8\mathrm{Li}(2^+)$
 & 1.32(6)    & \cdash      & 0.52(4)                  &  1.5(4)        &    0.6(2) & no & 0.014\\
$^9\mathrm{Li}(3/2^-_2)$\footnotemark[7] & $^8\mathrm{Li}(1^+)$
 & 0.34(6)    & \cdash      & 0.17(5)                   &  0.5(4)        &    $< 0.7$  & yes & 0.006\\
$^9\mathrm{Li}(3/2^-_2)$\footnotemark[7] & sum
 &        & 0.6(1) \cite{wuosmaa05-li9} & 0.69(6)\footnotemark[4]        &                    &   0.9(4)\footnotemark[4]  & \\\hline
$^9\mathrm{Be}(1/2^-)$ & $^8\mathrm{Be}(0^+)$
 & 1.11(12)  & 0.86(9)\footnotemark[8] \cite{chen70}   & 0.80(12)\footnotemark[4] &  4.4(6)       &    4.8(8)\footnotemark[4]  & yes &  0.0005\\
\hline
$^9\mathrm{Be}(7/2^-)$ & $^8\mathrm{Be}(0^+)$
 & 4.72(6)    & \cdash   &  0.0082(3)                        &   \cdash    &    \cdash   & yes & 0.005\\
$^9\mathrm{Be}(7/2^-)$ & $^8\mathrm{Be}(2^+)$
 & 1.69(6)    & \cdash   & 0.40(3)                        &   \cdash    &    \cdash   & yes & 0.003\\
$^9\mathrm{Be}(7/2^-)$ & sum
 &          &  1.2(2) \cite{dixit91}  & 0.41(3)\footnotemark[4]        &   \cdash    &    \cdash   & yes \\\hline
$^9\mathrm{B}(3/2^-)$ & $^8\mathrm{Be}(0^+)$
 & 0.185(1) \cite{audi03} & 0.00054(21) \cite{teranishi64} & 0.00058(2)\footnotemark[4] &  1.9(3)    &     0.9(3)\footnotemark[4]  & yes  & 0.0003\\
$^9\mathrm{B}(7/2^-)$ & $^8\mathrm{Be}(2^+)$
 & 4.13(6)  &  2.0(2) \cite{eval-A=8-10} & 0.82(4)\footnotemark[4]   &   \cdash    &    \cdash   & yes & 0.003\\
$^9\mathrm{C}(1/2^-)$ & $^8\mathrm{B}(2^+)$
 & 0.918(11) & 0.10(2) \cite{benenson74} & 0.102(5)    &  1.5(3)        &    0.43(26)  & yes & 0.006 \\
  \end{tabular}
\footnotetext[1]{I have computed an ``observed''
  width of $112\pm 15$ keV from the $R$-matrix formal width of 
  Ref.~\cite{bohlen01} and
  averaged it with the $160\pm 30$ keV FWHM of Ref.~\cite{stokes69}.}
\footnotetext[2]{From \cite{wuosmaa08}, based on ground state energy
from \cite{eval-A=5-7}.}
\footnotetext[3]{This is reported in Ref.~\cite{wuosmaa05-he7} as
  ``$\approx 2$ MeV'' with no quantitative error; a 1 MeV error is used in
  Figs.~\ref{fig:width-width}, \ref{fig:wigner}, and
  \ref{fig:woods-saxon} and in quoted statistics.}  
\footnotetext[4]{Open channels other than one-nucleon emission were
  neglected in the calculation (alpha or non-sequential).}
\footnotetext[5]{See Sec.~\ref{sec:isospin-mixing-be8} for discussion
  of the effects of isospin mixing in the observed $1^+$ and $3^+$
  states of $^8$Be.}  
\footnotetext[6]{Originally reported in
  Ref.~\cite{mcgrath67}, this number is construed in later work as
  the FWHM in the lab frame; since the center-of-mass excitation energy is
  reported in the same sentence, it appears to me to be a
  center-of-mass width.  Its error has apparently been mistranscribed
  in later references, independent of this ambiguity.}
\footnotetext[7]{Spin-parity assignments for $^9$Li follow 
  Ref.~\cite{wuosmaa05-li9}; see Sec.~\ref{sec:li9} below.}
\footnotetext[8]{The data compilations~\cite{eval-A=8-10} average the
  reported $R$-matrix formal width corresponding to this number with
  the much less certain Breit-Wigner width of Ref.~\cite{adloff71}.}
\end{table}

\begin{figure}
\includegraphics[width=3.5in,angle=0]{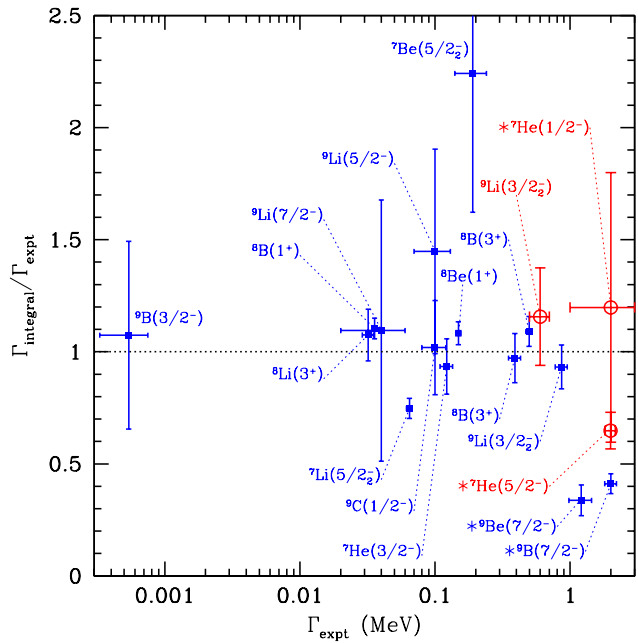}
  \caption{(Color online) The results of Table
    \ref{tab:big-width-table} for those cases included in the
    averages.  The vertical axis shows the ratio
    $\Gamma_\mathrm{integral}/\Gamma_\mathrm{expt}$ of the
    integral-relation width to the measured width, and the horizontal
    axis shows the measured width $\Gamma_\mathrm{expt}$.  The
    horizontal dotted line shows
    $\Gamma_\mathrm{integral}=\Gamma_\mathrm{expt}$ as a guide to the
    eye.  The errors shown are dominated by those on the experimental
    widths.  Experimental errors have been added in quadrature to
    smaller Monte Carlo statistical errors and errors propagated from
    the energy used in computing $\Gamma_\mathrm{integral}$.  States
    for which important alpha or 3-body decays have been neglected are
    indicated with asterisks next to the state labels, and large (red)
    circles indicate overlaps that seem inconsistent with $\pi/2$
    phase shift.  The outlying $5/2^-_2$ states of $^7$Be and $^7$Li
    are complicated by the need to include the small alpha decay
    channel in multichannel $R$-matrix fits when extracting widths
    from data; this introduces a considerable spread in reported
    nucleon widths \cite{eval-A=5-7}.}
  \label{fig:width-width}
\end{figure}

The computed widths of those states and channels for which
experimental data are available are shown graphically in
Fig.~\ref{fig:width-width}.  Only for those states where partial
widths are available from experiment do I show partial widths in the
graphs -- otherwise total widths are shown.
Fig.~\ref{fig:width-width} demonstrates the wide dynamic range of the
integral method, extending from 0.0005 to roughly 1 MeV.  With the
exception only of the very broad $^5\mathrm{He}$ states, which present
problems discussed below, all computed widths are within a factor of
three of experiment -- within a factor of two if states with
uncomputed alpha and direct three-body width are omitted.  The
error-weighted mean ratio
$\langle\Gamma_\mathrm{integral}/\Gamma_\mathrm{expt}\rangle$ of the
integral-method width to the experimental width is $0.82\pm 0.29$, and
the $\chi^2$ statistic for the difference between computed and
experimental widths is 5.9 per degree of freedom (omitting $^5$He for
a total of 19 states).  Restricting consideration to those states with
no omitted channels gives
$\langle\Gamma_\mathrm{integral}/\Gamma_\mathrm{expt}\rangle=1.09\pm
0.04$ and $\chi^2_\nu=1.5$ with $\nu=9$.  The errors used to compute
these statistics are mainly in the measured width
$\Gamma_\mathrm{expt}$.  

Setting aside the resonance energies, the main sources of uncertainty
in the theoretical calculation are in the accuracy of the variational
wave function and of the potential.  Because it is unclear how to
estimate these errors, the theoretical errors reported in
Table~\ref{tab:big-width-table} are the quadrature sums of Monte Carlo
sampling errors with errors propagated from the input resonance
energies.  Errors propagated from experimental energies are typically
small compared with the other errors.  However, each GFMC channel
energy is the difference of a resonance energy and a threshold energy
from separate calculations.  The error on this difference is often
large compared with the channel energy, and this propagates to a large
error on the predicted width.

\subsection{Correspondence of computed states to resonance states} 
\label{sec:psky}

The results presented in Table \ref{tab:big-width-table} and
Fig.~\ref{fig:width-width} vary in the degree to which they match the
experimental widths.  Some, like the first $3^+$ state of $^8$Li, are
close matches.  Others, like the two low-lying states of $^5$He, are
very far from agreement with experiment.  The $^5$He cases are
particularly interesting, because they are the only ones for which the
correct widths for the AV18+UIX Hamiltonian are known from explicit
scattering calculations \cite{NPWCH07}.  The known AV18+UIX widths are
1.5 MeV for the $3/2^-$ state and 5.0 MeV for the $1/2^-$ state, while
applying the integral relation to the pseudo\-bound VMC wave functions
gives 0.68 and 0.71 MeV, respectively.  The $1/2^-$ state lies above
the centrifugal barrier and is so broad that its phase shift does not
pass through $\pi/2$, so it is unsurprising that the integral method
fails for this case.  The $3/2^-$ state does not present these
difficulties, so its difference from both experiment and theory is
clearly not a shortcoming of the potential but rather of the
computational methods.

There are several reasons that the application of the integral
relation to a pseudo\-bound variational wave function could fail to
yield the correct width for the potential.  I begin by noting that for
the states examined, the function $F_{l_a}\!(\eta_a,k_ar_a)$ typically
does not deviate far from its leading-order dependence on $k_ar_a$
over the range of $r_a$ where $\urel^{a,p} - V^a_C$ differs
significantly from zero.  The main effect of changing the assumed
resonance energy is therefore to change the overall scale of the
integrand in Eq.~(\ref{eq:g-norm-integral}) without changing its shape
much.  Thus, simply using the experimental channel energy in the
integral relation should correct rather accurately for the mismatch
between experimental and theoretical channel energies without
introducing significant distortions in the integrand.

It is possible that the variational minimization with constrained
charge radius fails to produce good approximations to some resonance
states.  It could happen that the variational ansatz is a poor match
to a particular resonant wave function or that the VMC wave function,
being only approximate, contains ``contamination'' from nonresonant
continuum states.  Contamination more energetic than the desired state
is precisely what the GFMC method is intended to remove, and much of
the success of the QMC methods lies in the exclusion of low-energy
excitations from the variational wave function.  Any off-resonance
contamination produces contributions to the integral relation that do
not correspond to the pole residue.  The danger of contamination would
seem to grow with the resonance width, because a less-peaked density
of states implies a wider range of states that are similar to the
resonance and difficult to eliminate by variational minimization.  The
presence of nearby resonances in the same channel may compound this
problem by contributing contamination with rather different structure
from the resonant wave function being sought.

Yet another kind of difficulty lies in the normalization of the wave
function.  The relationship between the width and the asymptotic
normalization of standing-wave states found in
Refs.~\cite{humblet61,humblet70,humblet90} depends on the wave
function being normalized in a finite volume and its amplitude being
small at the boundary of that finite volume as expressed in
Eq.~(\ref{eq:humblet-correction}).  I have assumed that the
normalization volume is effectively defined by the exponential
fall-off of the $f^{ss}$ and $\phi_p^{LS[n]}$ correlations in the
tails of the variational wave function.  If either the true or the
computed wave function fails to fall off rapidly enough, the
normalization is problematic in ways that can be viewed either as the
lack of an effective cutoff or as neglect of a large surface-amplitude
correction.

The normalization problem is perhaps the most straightforward to
examine.  The final column of Table \ref{tab:big-width-table} 
displays a parameter
\begin{equation}
\label{eq:zeta}
  \zeta = \left(\frac{8\ \mathrm{fm}\ \times\ R_a(8\ \mathrm{fm})}{
    r_\mathrm{max} R_a(r_\mathrm{max})}\right)^2,
\end{equation}
where $r_\mathrm{max}$ is the location of the maximum $|R_a(r)|$, and
$R_a(r)$ is computed directly from Eq.~(\ref{eq:direct-overlap}).
This ratio measures the amplitude of the variational wave function
just outside the interaction region relative to that in the interior
(accounting for the $r^2$ dependence of the volume of a spherical
shell).  We may expect to encounter difficulties when $\zeta \gtrsim
0.1$, and this occurs for four states: the two $^5$He states and the
$1/2^-$ and $5/2^-$ states of $^7$He.  The relevance of $\zeta$ may
also be viewed in light of Eq.~(\ref{eq:humblet-correction}).  If I
adopt 8 fm as the effective boundary of the normalization volume and
assume (cf.  Fig. \ref{fig:psky} below) that typically
$R_a(8\ \mathrm{fm})\sim 0.02\ \mathrm{fm}^{-3/2}$, then
Eq.~(\ref{eq:humblet-correction}) gives $\epsilon < 0.2$ for all but a
very few of the widths computed here.  Much larger values of
$\epsilon$ occur for $^5\mathrm{He}(1/2^-)$, $^7\mathrm{He}(1/2^-)$,
and minor decay channels of a few other states.

The overlap functions may also be tested for consistency with
expectations for a resonance.  The integral relation for the width is
the $r\rightarrow\infty$ limit of $R_a$ as computed from
Eq.~(\ref{eq:full-integral-relation}).  Given a wave function, it is
possible to compute the overlap function at all radii using
Eq.~(\ref{eq:full-integral-relation}) given $B_{a,\infty}$, and the
result is likely more accurate than that from direct calculation of
$R_a$.

Determining $B_{a,\infty}$ from a variational wave function is a
tractable problem \cite{kievsky10,romero-redondo11}, but I do not
pursue it here.  Since I have tacitly \textit{assumed} in computing
widths that $B_{a,\infty}=0$ as required for a resonance state
(cf.~Eq.~(\ref{eq:kmatrix-integral})), I can compute the full overlap
function from the integral relation
(Eq.~(\ref{eq:full-integral-relation})) with $B_{a,\infty}=0$ and
check that it matches the overlap function computed directly from
(Eq.~(\ref{eq:direct-overlap})).  If the two overlaps are in rough
agreement in the interaction region, then the VMC wave function is
consistent with a resonance wave function and may validly be used to
compute a width.

\begin{figure}
\includegraphics[width=5.5in,angle=270]{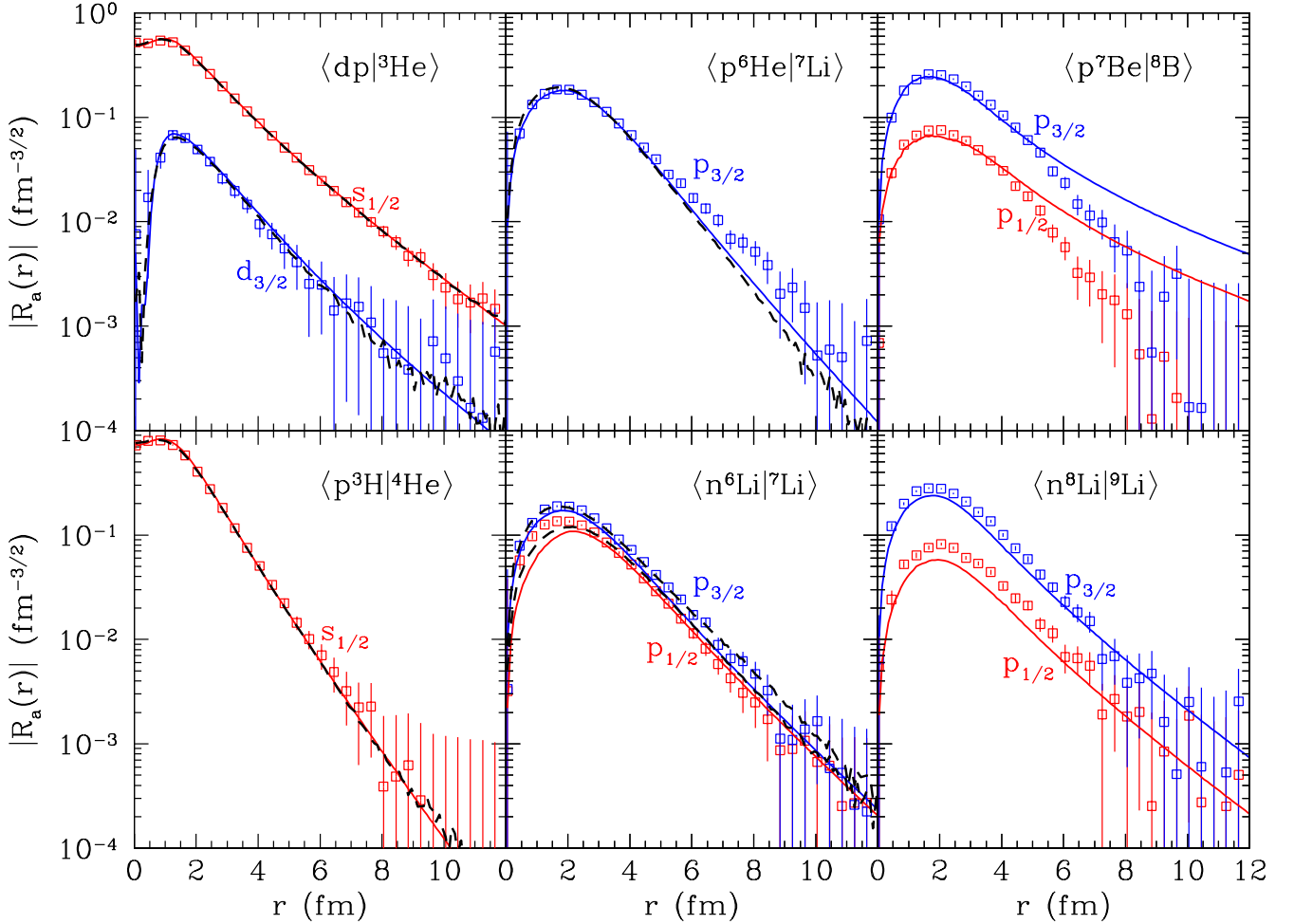}
  \caption{(Color online) Overlap functions for several bound states,
    computed from the bound-state analogue of the integral relation in
    Eq.~(\ref{eq:full-integral-relation}) (solid curves) and from the
    definition in Eq.~(\ref{eq:direct-overlap}) (squares with Monte
    Carlo error bars.)  The $l_a$ and $j_a$ quantum numbers of the
    ``virtually emitted'' nucleon are indicated by labels near the
    appropriate curves (and distinguished by color).  Where they
    exist, overlaps computed directly from GFMC wave
    functions~\cite{brida11} are shown as (black) dashed curves.}
  \label{fig:bound-overlaps}
\end{figure}

As a test of the approach, I apply it to bound states, since their
integration constant corresponding to $B_{a,\infty}$ is zero by
definition.  Results are shown for several states in
Fig.~\ref{fig:bound-overlaps}, and I have computed them for all of the
channels considered in Ref.~\cite{nw11}.  The agreement between the
two calculations, especially for $s$-shell nuclei where the VMC method
is more accurate, is excellent.  The deviations of the
integral-relation overlap from the direct overlap are likely to be
improvements: the integral relation contains more information about
the potential than does the VMC wave function by itself, and it
guarantees the correct $r\rightarrow\infty$ asymptotics.  For some
nuclei with $A=3$, 4, and 7, GFMC calculations of overlaps exist
(albeit for the AV18+IL7 Hamiltonian, not AV18+UIX) \cite{brida11}.
The results of the GFMC calculations (dashed curves of
Fig.~\ref{fig:bound-overlaps}) are not severely different from those
of applying the integral method to VMC wave functions, and they
deviate from the VMC overlaps by similar amounts.  This experience
with bound states indicates that overlaps computed from integral
relations are at least as accurate as directly-computed overlaps of
VMC wave functions and are not in conflict with GFMC results.  They
may therefore be very useful for calculations of spectroscopic factors
and transfer and knockout cross sections.

Overlap functions are shown for $^8\mathrm{B}(1^+)$ in
Fig.~\ref{fig:fixing-asymptotics} and for several other resonance
states in Fig.~\ref{fig:psky}.  These were computed both directly and
by integral relations with $B_{a,\infty}=0$.  In all cases, the
integral relation replaces the artificial exponential fall-off of the
VMC wave function with the oscillatory behavior of $G_{l_a}$.  Based
on experience with bound states, the two overlaps should agree at
$r\lesssim 4$ fm.  In some cases (e.g.~$^9\mathrm{C}(1/2^-)$) the
match there is quite good, while in others (e.g.~$\langle
n\,^6\mathrm{He}|^7\mathrm{He}(1/2^-)\rangle$) overlaps from the two
methods seem to have little to do with each other.  I conclude that
when results of the two methods are qualitatively very different
inside 4 fm, there is an inconsistency with the assumption of $\pi/2$
phase shift so that the pseudo\-bound VMC wave function may not allow
accurate width calculations.  The penultimate column of Table
\ref{tab:big-width-table} indicates for each width a qualitative
judgment of whether the two methods of computing overlaps agree, and
cases where they do not are indicated in Fig.~\ref{fig:width-width} by
large (red) circles.  Examination of the table reveals that with the
exception of the $0^+$ state in $^8$B, failures of the ``${\pi}/{2}$
assumption'' occur only in a few neutron emission channels and in no
proton emission channels.  Considering only states consistent with
$\delta_l=\pi/2$, I find
$\langle\Gamma_\mathrm{integral}/\Gamma_\mathrm{expt}\rangle= 0.83\pm
0.30$ and $\chi^2_\nu=6.3$, this time for $\nu=15$ instead of 18.
Further restriction to states in which all decay channels are computed
gives $\langle\Gamma_\mathrm{integral}/\Gamma_\mathrm{expt}\rangle =
1.08 \pm 0.04$ and $\chi^2_\nu = 1.6$ for 8 degrees of freedom.  These
are essentially the same results as for the entire data set.  It
appears that the best predictor of whether a calculated width will
match experiment is simply whether the calculation includes all
channels contributing to the measured width.

\begin{figure}
\includegraphics[width=5.5in,angle=270]{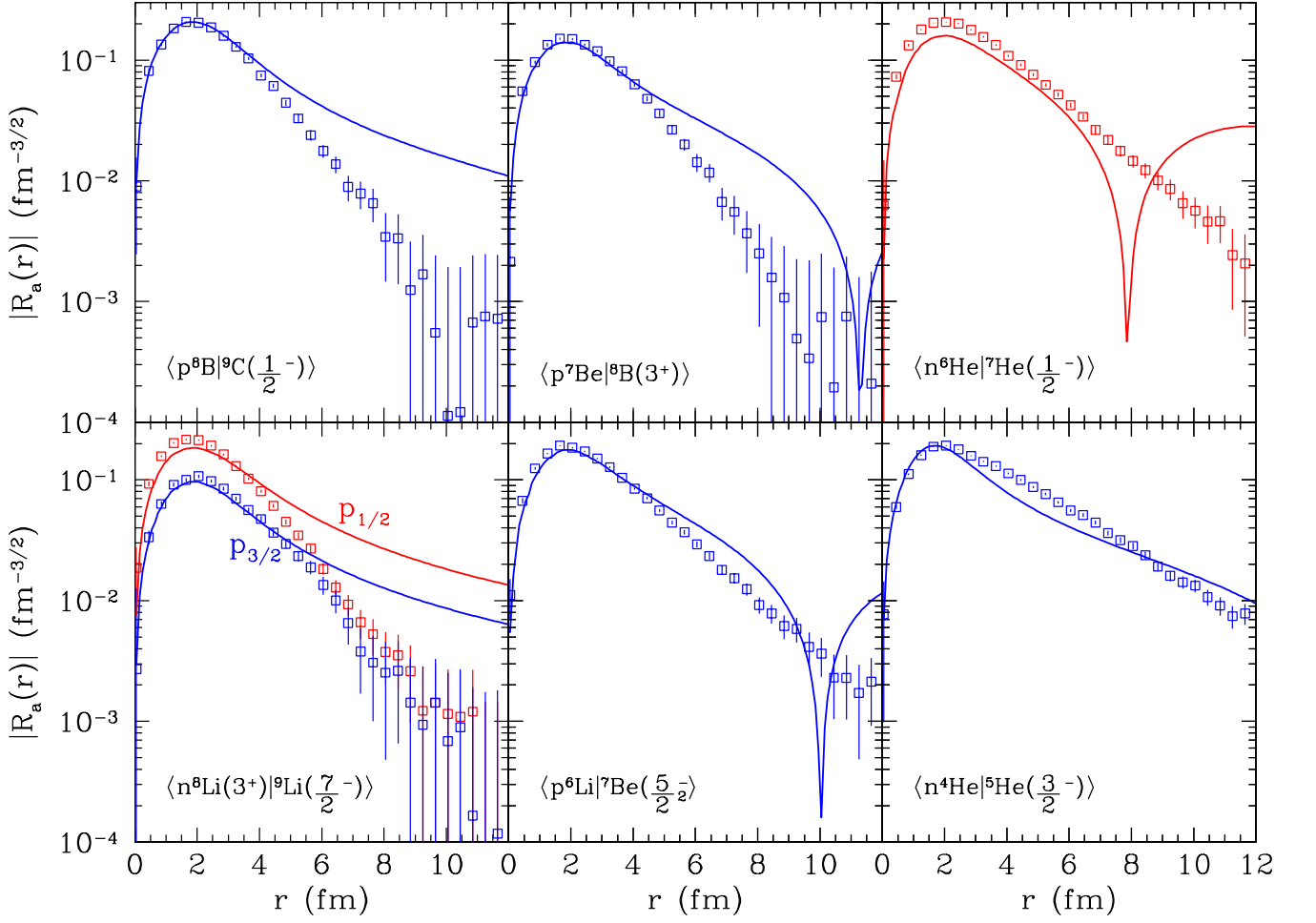}
  \caption{(Color online) Overlap functions of several resonance
    states, computed directly from the definition
    (Eq.~(\ref{eq:direct-overlap}); points with Monte Carlo errors)
    and from the integral relation of
    Eq.~(\ref{eq:full-integral-relation}) with $B_{a,\infty}=0$ (or
    $\delta_{l_a}={\pi}/{2}$; solid curves).  Where more than one
    angular momentum channel is available, the channels are labeled.
    The four cases on the left show good agreement between the two
    methods at small separation ($r\lesssim 4$ fm), and the
    corresponding widths agree with experiment.  The two states on the
    right are inconsistent with the $\delta_{l_a}=\pi/2$ assumption
    since the two types of overlap calculations disagree at $r< 4$ fm.
    Computed widths of these two states also disagree with
    experiment.}
  \label{fig:psky}
\end{figure}

\subsection{Isospin mixing in $^8\mathrm{Be}$}
\label{sec:isospin-mixing-be8}

A difficulty presents itself in considering the pair of $J^\pi=3^+$
states of $^8$Be at $\sim 19$ MeV and the pair of $1^+$ states at
$\sim 18$ MeV.  Each of these doublets consists of one $T=0$ and one
$T=1$ state that have mixed, with the lower state of each pair
predominantly $T=1$.  The VMC wave functions have definite isospin, so
incorporating the mixing into calculations of their widths presents
ambiguities.  Simply ignoring the mixing and assigning the $T=1$
widths to the lower states and $T=0$ widths to the higher states does
not work.  For example, in the $3^+$ doublet, the lower state has
measured width $270\pm 20$ keV and the upper has $227\pm 20$ keV; the
unmixed widths from the integral method are 177 keV and 364 keV,
respectively.

There exist preliminary GFMC calculations of the mixing matrix element
of the Hamiltonian for each of these doublets \cite{wiringa-private},
and this matrix element can be combined with the splitting of the
doublet to determine its mixing angle (e.g. \cite{barker-ferdous78}).
The splittings of the doublets are small enough that they are not
resolved by the existing GFMC calculations, so the experimental
splittings must be used to compute mixing angles.  Since the
recommended \cite{eval-A=8-10} 165 keV splitting of the $3^+$ doublet
is less than twice the GFMC mixing energy, these two numbers cannot be
combined to yield a real mixing angle.  (Adopting a splitting of 310
keV \cite{barker78} produces good agreement with the experimental
widths.)  In the $1^+$ doublet, combining the GFMC mixing matrix
element with the experimental splitting produces a poor match to the
data (widths of 14.2 and 105 keV \textit{versus} $10.7\pm 0.5$ and
$138\pm 6$ keV from experiment).

For each doublet, a mixing angle can be computed by $\chi^2$
minimization of the difference between theoretical and experimental
widths.  This produces good agreement with the measured widths, which
was not guaranteed.  The resulting mixing angle for the $3^+$ doublet
has $\sin\theta_\mathrm{mix}=0.28$, which is small relative to
literature values \cite{barker78}.  The mixing angle from minimizing
$\chi^2$ for the $1^+$ doublet has $\sin\theta_\mathrm{mix}=0.068$,
much smaller than both the value of $0.20$ from the GFMC mixing energy
and the literature value of $0.21$ \cite{oothoudt77}.

Thus, attempts to compute separate widths for the upper and lower
states in each doublet run aground on the problem of finding mixing
angles consistent with all available information.  For each doublet,
the quantity most independent of the mixing angle is the sum of the
two widths.  The sum is less sensitive to $\theta_\mathrm{mix}$ than
are the underlying pole residues, but it is not quite independent of
$\theta_\mathrm{mix}$ because the doublet states are not degenerate.
It is the sum for each doublet that is shown in Table
\ref{tab:big-width-table} and Fig.~\ref{fig:width-width}, with the
$T=1$ width computed in each case using the lower energy and the $T=0$
width the higher.  These sums are in good agreement with experiment.

\subsection{Applications to recent measurements}

An important use of theoretical estimates of widths is in the
identification of observed states.  Most of the states considered here
($A\leq 9$, $\Gamma \lesssim 1$ MeV, dominated by nucleon decays) were
found experimentally, and their spins and parities identified, long
ago.  Some exceptions are the first $0^+$ and second $2^+$ states of
$^8$B and the entire low-lying spectra of $^9$He and $^9$Li.  Here I
attempt to shed light on these systems by calculating widths from VMC
wave functions.

\subsubsection{$^8\mathrm{B}$}

Several theoretical models predict low-lying states of $^8$B that have
not been observed, as discussed in Ref.~\cite{mitchell10}.  It is
possible that the states are simply too broad to be seen easily in
experiments, and evidence was recently found for a $0^+$ state at 1.9
MeV above the $^8$B ground state \cite{mitchell10}.  An $R$-matrix fit
to both elastic and inelastic scattering of $^7$Be on protons
indicated a $0^+$ state with partial width $0.28\pm 0.14$ MeV for
decay to the $^7$Be ground state and $0.33\pm 0.18$ MeV to the first
excited state.  Since these widths are within $2\sigma$ of zero, I
have not shown them in Table \ref{tab:big-width-table}.  This state
has some support from the calculation of Ref.~\cite{navratil11-be7pg},
where it was found in computed phase shifts at the same energy using
the merged no-core shell model and resonating group method.  That
calculation also indicates reason for caution in applying the integral
relation: the phase shift does not approach $\pi/2$, and indeed the
VMC $0^+$ state fails the $\pi/2$ consistency test discussed above.

\begin{figure}
\includegraphics[width=3in]{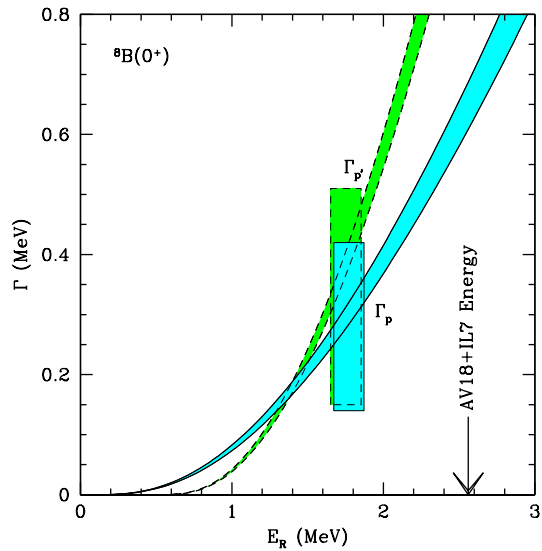}  
  \caption{(Color online) Predicted partial widths as functions of
    assumed resonance energy (relative to the $^7$Be ground state
    threshold) for the first $0^+$ state in $^8$B.  The thicknesses of
    the bands indicate the errors from Monte Carlo sampling.  The band
    bounded by solid curves indicates the partial width to the ground
    state of $^7$Be, and that bounded by dashed curves shows the
    partial width to the first excited state.  The boxes (offset from
    each other slightly in the horizontal direction for visibility)
    show the experimental results of Ref. \cite{mitchell10} without a
    correction from $R$-matrix formal width to observed width. (This
    can amount to a $\sim 30\%$ reduction at the high end of the
    allowed width.)  Also indicated is the best estimate of the
    resonance energy from GFMC calculations using the AV18+IL7
    Hamiltonian.}
  \label{fig:b8_0}
\end{figure}

I show in Table \ref{tab:big-width-table} the predicted partial widths
of the $0^+$ state at its energies for the AV18+UIX and AV18+IL7
Hamiltonians.  Fig.~\ref{fig:b8_0} shows the dependence of this
prediction on the assumed resonance energy.  The claimed experimental
widths are consistent with my results, though inconsistency of the
overlaps with ${\pi}/{2}$ phase shift makes the significance of this
consistency doubtful.  I also include in Table
\ref{tab:big-width-table} the AV18+IL7 energies and VMC-computed
partial widths of the unobserved isobaric-analogue state in $^8$Li.
The $^8$Li and $^7$Li VMC wave functions are explicitly isospin
rotations of the $^8$B and $^7$Be wave functions, and the
$^8\mathrm{Li}(0^+)$ decays also fail the $\pi/2$ condition on the
overlaps.

The authors of Ref.~\cite{mitchell10} also find a $2^+$ state at 2.55
MeV excitation.  Because this result has not been confirmed, I omit
this state from goodness-of-fit statistics, but I have computed its
partial widths to the $^7$Be ground and first-excited states and
included the results in Table \ref{tab:big-width-table}.  The overlap
functions for these states are compatible with $\pi/2$ phase shift.
However, the computed partial widths do not match those claimed in
Ref.~\cite{mitchell10}: I find $0.51\pm 0.02$ MeV to the $^7$Be
ground state \textit{versus} $0.12\pm 0.04$ MeV measured, and
$0.039\pm 0.002$ MeV to the $^7$Be excited state \textit{versus}
$0.24\pm 0.11$ MeV measured.  The origin of these differences is not
clear.

\subsubsection{$^9\mathrm{He}$}

The spectroscopy of $^9$He remains unclear despite several
experimental studies
\cite{seth87,bohlen99,chen01,rogachev03,barker04,volya05,golovkov07,johansson10}.
The ground state was originally thought to be a $1/2^-$ resonance
state \cite{seth87}.  Subsequently, strong $s$-wave $n$-$^8$He
interaction was seen near threshold and argued to reflect a $1/2^+$
virtual state~\cite{chen01,bohlen99}.  More recent experiments have
found a smaller scattering length and thus less support for a virtual
state~\cite{golovkov07,johansson10}.  Because $s$-wave neutrons do not
have true resonances passing through $\pi/2$ phase shift, I do not
present a width for this state.

Several other observations of resonances within a few MeV of the
$^8$He+$n$ ``threshold'' have been claimed.  (See
Ref. \cite{johansson10} for a summary.)  Spin and parity assignments
for all of these states are uncertain, and matching them to
theoretical expectations has proven difficult
\cite{chen01,eval-A=8-10,barker04}.  Width estimates based on
\textit{ab initio} calculations could provide useful guidance, so I
explore this possibility here.

There have been four claims of a state around 1.2 MeV above the
$^8$He+$n$ ``threshold''
\cite{seth87,bohlen99,johansson10,belozerov88}, and it has additional
support from a study of possible analogue states \cite{rogachev03}.
This is generally assumed to be the lowest-lying $p$-shell state, with
$J^\pi=1/2^-$, and there is conflicting information concerning its
width.  Ref.~\cite{seth87} found it to be narrower than their 0.42 MeV
resolution; other experiments found 1 MeV \cite{belozerov88}, $0.10
\pm 0.06$ MeV \cite{bohlen99}, and 2 MeV \cite{golovkov07}.  As
pointed out particularly by Barker \cite{barker04}, it is difficult to
reconcile widths considerably narrower than 1 MeV with the expected
strong single-particle character of the $1/2^-$ resonance.
Theoretical calculations also place the $1/2^-$ state a few MeV higher
(e.g. at 3 and 4 MeV in GFMC calculations with Illinois-6 and
Illinois-2 three-body forces, respectively).  One possibility is
reduction of the $1/2^-$ energy by an $sd$-shell component that is
missing in the calculations~\cite{PVW02}.

Additional resonances have been found at higher energies, one around
2.3 MeV \cite{seth87,bohlen99,johansson10} with claimed width $0.7\pm
0.2$ MeV, one around 4 MeV \cite{bohlen99,golovkov07}, and another
around 5 MeV \cite{seth87,bohlen99}.  In addition to the $1/2^-$
state, a $3/2^-$ state is expected theoretically, though also at
higher energy.  It is also likely that $sd$-shell intruder states with
$J^\pi=5/2^+$ and $3/2^+$ are present in the low-lying spectrum.

\begin{figure}
\includegraphics[width=3in,angle=0]{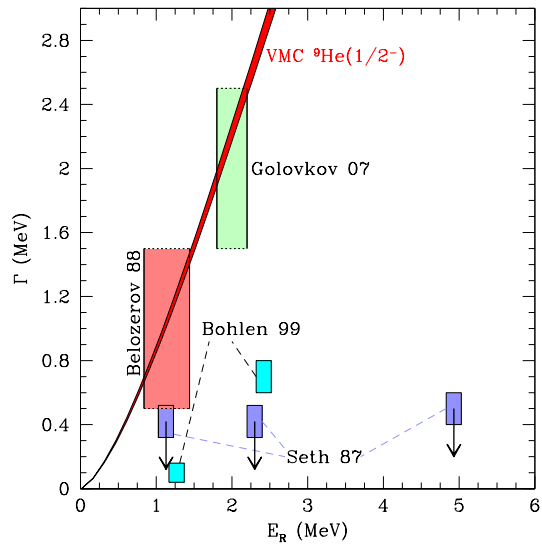}
  \caption{(Color online) Boxes (with sizes indicating reported
    errors) show claimed $^9$He resonances that have reported
    widths.  The band labeled ``VMC'' shows the predicted width of the
    $1/2^-$ state as a function of the resonance energy assumed in the
    integral relation, and its width reflects the statistical error of
    the Monte Carlo integration.  Where no error was reported for a
    width (for example, Ref. \cite{golovkov07} has ``$\sim 2$ MeV''
    for the width of the state at 2.0 MeV), I have assigned an error
    of 0.5 MeV and indicated this with broken lines at the upper and
    lower limits.  I have omitted some very broad states with missing
    or lower-limit errors: a state with $E_R\sim 3$ MeV and
    $\Gamma\sim 3$ MeV from Ref. \cite{belozerov88} and one with
    $E_R\geq 4.2$ MeV and $\Gamma>0.5$ MeV from
    Ref. \cite{golovkov07}.  See Ref. \cite{johansson10} for
    additional reports of states without measured widths.  The
    displayed widths are from Seth \textit{et al.}  \cite{seth87},
    Belozerov \textit{et al.}  \cite{belozerov88}, Bohlen \textit{et
      al.} \cite{bohlen99}, and Golovkov \textit{et al.}
    \cite{golovkov07}.  Downward arrows indicate widths that include
    $\sim 400$ keV instrumental resolution.}
  \label{fig:9he}
\end{figure}

I computed widths of $p$-shell states with $J^\pi=1/2^-$ and $3/2^-$,
but the results proved difficult to match with experimental data.  The
short-range overlaps of the $1/2^-$ state computed directly and by the
integral method are in nice agreement.  Since it is unclear what
resonance energy should be used in the calculation, I have computed
widths from the VMC wave function using a range of channel energies in
the integral relation.  Fig.~\ref{fig:9he} shows these results.  They
mainly demonstrate the argument of Barker \cite{barker04} that there
is a mismatch between the narrow width of the $1/2^-$ state claimed by
Bohlen \textit{et al.}~\cite{bohlen99} and theoretical expectations of
a strongly single-particle state.  The resonances claimed by Belozerov
\textit{et al.}~\cite{belozerov88} and Golovkov \textit{et
  al.}~\cite{golovkov07} are consistent with the computed $1/2^-$
width, but that assignment makes interpretation of the 1.3 and 2.4 MeV
states of Bohlen \textit{et al.}~\cite{bohlen99} difficult: at least
one would have to be an $sd$-shell state.

All channel energies below 6 MeV for the $3/2^-$ state yield computed
widths of less than 5 keV.  This could match either of the states at
4.3 and 5.3 MeV in Ref. \cite{bohlen99}, which were found to be
narrower than the 100-keV experimental resolution, but the VMC
overlaps of this state are inconsistent with $\pi/2$ phase shift.  For
this reason, the integral-relation results are probably not reliable
predictions of the width.

VMC wave functions also exist for states with $J^\pi=1/2^+, 3/2^+,$
and $5/2^+$, but these wave functions with $sd$-shell components have
not reached the same level of development as the VMC $p$-shell states
and have not been published.  As mentioned above, the $1/2^+$ state
should not be observable as a resonance.  The VMC wave functions for
the $3/2^+$ and $5/2^+$ states indicate that they are mainly made by
coupling the $2^+$ state of $^8$He to an $s$-wave neutron and
therefore also not true $\delta_a=\pi/2$ resonances.  I did not
attempt calculations of partial widths for decay to the
$^8\mathrm{He}(2^+)$ state.  Calculations of the partial widths for
decay from positive-parity states to the $^8$He ground state by
emission of a $d$-wave neutron yielded partial widths of less than 60
keV and overlaps inconsistent with $\pi/2$ phase shift.

\subsubsection{$^9\mathrm{Li}$}
\label{sec:li9}

The most recent data compilation for $A=9$ lists five low-lying states
of $^9$Li and only assigns firm spins and parities to the lowest two
\cite{eval-A=8-10}.  A more recent paper \cite{wuosmaa05-li9}
identifies the third state as $J^\pi=5/2^-$ by comparing spectroscopic
factors of VMC wave functions (older versions of those used here) with
measured $(d,p)$ cross sections.  Those authors also assigned
$J^\pi=3/2^-$ and $7/2^-$, respectively, to the next two states.  This
assignment was based partly on the ordering of states in theoretical
calculations and partly on the assumption that widths should correlate
with computed spectroscopic factors.  The results presented in Table
\ref{tab:big-width-table} and Fig.~\ref{fig:width-width} support these
assignments by reproducing the widths of all three unbound states.

\subsection{Comparison with other width estimates}
\label{sec:comparison}

I conclude by considering other ways to estimate widths from VMC wave
functions and comparing them with the integral method.  In the absence
of integral relations or explicit scattering calculations, widths must
be estimated from spectroscopic factors.  In applications of the shell
model, one often assumes that the width is the product of the
spectroscopic factor,
\begin{equation}
\label{eq:spectroscopic-factor}  
 S_a \equiv \int_0^\infty \left[R_a(r)\right]^2r^2\,dr\,,
\end{equation}
and the single-particle width.  VMC spectroscopic factors might in
fact be more suited to this procedure than those from a shell model,
because shell models are typically confined to a single value of the
principal quantum number, while the large amount of correlation in VMC
wave functions guarantees that Eq.~(\ref{eq:direct-overlap}) picks up
contributions from all major shells.

The crudest estimate of the single-particle width is the ``Wigner
limit'' \cite{teichmann52}.  On the basis of a causality argument, the
width of a resonance can be shown to have an approximate upper limit
of
\begin{equation}
\label{eq:wigner}
  \Gamma_W = 2 \frac{\hbar^2}{\mu_a b^2}P_l(k_a b),
\end{equation}
where $b$ is a ``matching radius'' (typically $\sim 4$ fm) defining
the edge of the interaction region, and $P_l$ is the penetration
factor of Ref.~\cite{lanethomas}.  Since $\Gamma_W$ is (approximately)
an upper limit on the width that a resonance can have, it might
approximate a single-particle width.  Some authors define $\Gamma_W$
to include an additional numerical factor multiplying
Eq.~(\ref{eq:wigner}), reflecting assumptions about the wave function
inside the interaction region.  Teichmann and
Wigner~\cite{teichmann52} assumed a constant wave function to arrive
at a factor of $3/2$.  Other authors make more elaborate assumptions
and arrive at a factor of $(2l-1)/(2l+1)$ for $l\neq 0$
\cite{bohrmottelson}.  I take Eq.~(\ref{eq:wigner}) to define
$\Gamma_W$.

\begin{figure}
\includegraphics[width=3.5in,angle=0]{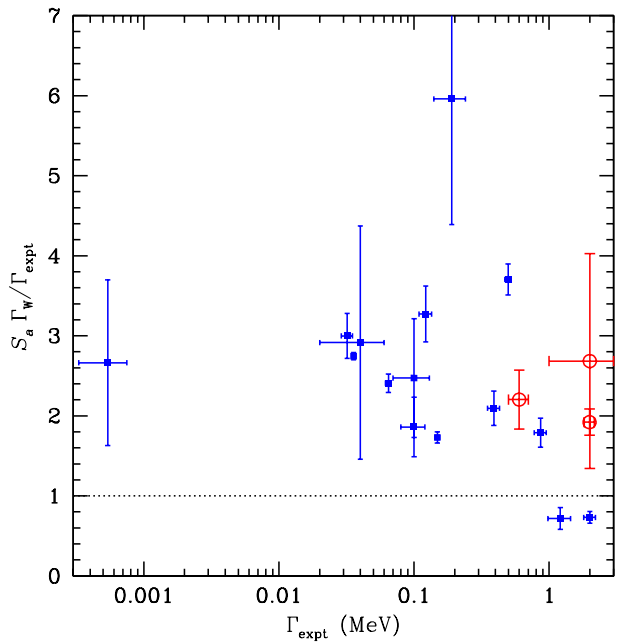}
  \caption{(Color online) Comparison between experimental widths and
    widths estimated as the VMC spectroscopic factor times the Wigner
    limit.  The vertical axis shows the ratio of these numbers, using
    a radius of 4 fm to compute the Wigner limit.  Symbols are as in
    Fig.~\ref{fig:width-width}, and the labels from that figure may be
    used to identify states here.  Comparison with
    Fig.~\ref{fig:width-width} indicates that the integral relation is
    a significantly better predictor of widths than $S_a\Gamma_W$.
    Note the different vertical scale from
    Fig.~\ref{fig:width-width}.}
  \label{fig:wigner}
\end{figure}

In Fig.~\ref{fig:wigner} I use the Wigner limit to estimate widths of
the states under consideration.  In each case, I multiply $\Gamma_W$
for a 4 fm radius by the VMC spectroscopic factor from
Eqs.~(\ref{eq:direct-overlap}) and (\ref{eq:spectroscopic-factor}).
These estimates plainly do not reproduce measured widths as well as
the integral relation.  The weighted mean ratio of ``theoretical'' to
experimental width for this method, restricted to states consistent
with $\pi/2$ phase shift and purely nucleon-emission decay, is
$2.49\pm 0.52$; the reduced $\chi^2$ is 1845 for eight degrees of
freedom.  (Recall that
$\langle\Gamma_\mathrm{integral}/\Gamma_\mathrm{expt}\rangle = 1.08\pm
0.04$ with $\chi_\nu^2=1.6$ for the same set of states.)  The mismatch
between $S_a\Gamma_W$ and $\Gamma_\mathrm{expt}$ can be reduced by
choosing a smaller numerical factor to define the Wigner limit, but
that does not remove the large scatter in
$S_a\Gamma_W/\Gamma_\mathrm{expt}$.  Better agreement with experiment
is likewise achieved with a smaller radius, but good agreement
requires an unphysically small radius in the neighborhood of 2 fm,
which again does not remove the large scatter in the ratio
$S_a\Gamma_W/\Gamma_\mathrm{exp}$.  Estimates from $\Gamma_W$ are
typically $\sim 20\%$ smaller if they are estimated as ``observed
widths'' from Eq.~(\ref{eq:observed-width}), using
\begin{equation}
  \gamma_a^2 = {S}_a\frac{\hbar^2}{\mu_a b^2}\ .
\end{equation}
This helps significantly with neither the overall scale nor the large
scatter of the predicted widths.

A better estimate of the single-particle width, and one perhaps more
widespread in shell-model studies, is based on Woods-Saxon potentials.
One assumes a potential well of ``standard'' radius and diffuseness
and adjusts its depth to produce a resonance at the correct energy.
The width $\Gamma_{WS}$ of this resonance is then taken as an estimate
of the single-particle width.  A range of geometric parameters for the
potential is usually considered, because the most appropriate values
are not known \textit{a priori}.

\begin{figure}
\includegraphics[width=3.5in,angle=0]{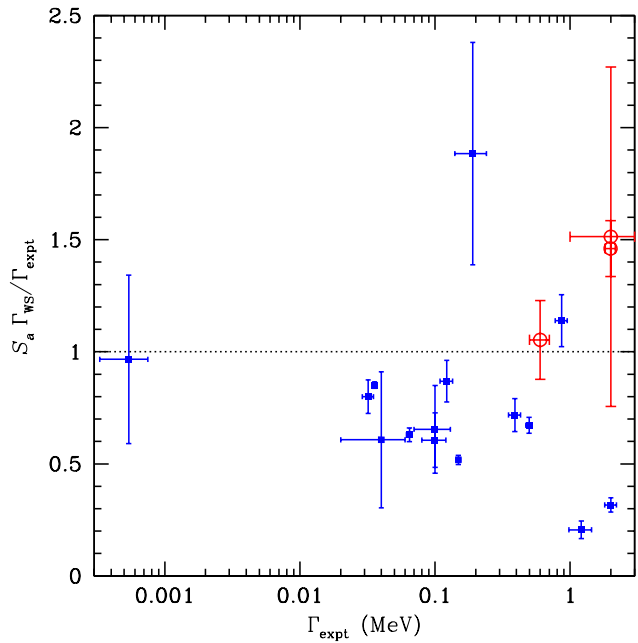}
  \caption{(Color online) Widths estimated as the VMC spectroscopic
    factor times the Woods-Saxon single-particle width
    $\Gamma_\mathrm{WS}$, divided by experimental widths.  Symbols are
    as in Fig.~\ref{fig:width-width}, and the labels from that figure
    may be used to identify states here.  Comparison with
    Figs.~\ref{fig:width-width} and \ref{fig:wigner} indicates that
    this approach predicts experimental widths more accurately than
    the Wigner-limit approach but still not as well as the integral
    relation.}
  \label{fig:woods-saxon}
\end{figure}

I computed estimates of this kind, using diffuseness 0.65 fm and
Woods-Saxon radius $1.1\ \mathrm{fm}\ \times (A-1)^{1/3}$, with $A$
the mass number of the resonance state (the defaults in a code
provided by B.~A.~Brown \cite{brown-private}).  I neglected variation
of these parameters and estimated the single-particle width
$\Gamma_\mathrm{WS}$ from the FWHM of the peak in the computed cross
section for the given Woods-Saxon well.  The comparison of
$S_a\Gamma_\mathrm{WS}$ with $\Gamma_\mathrm{expt}$ is shown in
Fig.~\ref{fig:woods-saxon}.  The Woods-Saxon estimates are
systematically low, with a weighted mean $\langle
S_a\Gamma_\mathrm{WS}/\Gamma_\mathrm{expt}\rangle= 0.74 \pm 0.15$ for
the same eight cases considered above and $\chi^2_\nu= 34$ for eight
degrees of freedom.  I conclude from this exercise and the similar
exercise using Wigner limits that VMC widths computed by the integral
method are more useful predictors of experimental widths than are the
VMC spectroscopic factors.

\section{Summary}
\label{sec:conclusion}

I have presented plausibility arguments, supported by detailed
derivations in the literature, that widths of resonant states can be
estimated by evaluating an integral over pseudo\-bound \textit{ab
  initio} wave functions.  This approach is approximate, but it avoids
a great deal of computation and human labor that would be needed in
full-on scattering calculations and would often be complicated by
coupled channels.  It is nicely suited to quantum Monte Carlo
calculations in that it is insensitive to the difficult-to-compute
tails of the many-body wave functions, it involves a short-range
integral amenable to Monte Carlo integration, it uses more information
about the Hamiltonian than is encoded in the variational wave
function, and it can be applied to resonances narrower than the
practical energy resolution of the GFMC technique.  Related integrals
yield overlap functions for bound states, and these overlaps are
guaranteed to have the correct shapes in their long-range asymptotics
even when the variational wave function does not.  These may be useful
for calculations of spectroscopic factors and of transfer and knockout
cross sections.

I have implemented integral-method width calculations for one-nucleon
emission from wave functions computed by the variational Monte Carlo
method.  It yields widths in good agreement with experiment for
several states in the $7\leq A \leq 9$ mass range.  Cases of
disagreement always involve either open channels for which I have not
accounted or a resonant wave function that is not strongly peaked in
the interaction region.  I have shown that widths predicted in this
way are closer matches to experiment than are na\"ive combinations of
\textit{ab initio} spectroscopic factors with Wigner-limit or
Woods-Saxon estimates of single-particle widths.  The integral method
is thus a useful tool for estimating widths from \textit{ab initio}
methods that produce pseudo\-bound wave functions.

For the longer term, the calculations presented here represent a
learning problem for application of integrals of the type in
Eqs.~(\ref{eq:f-norm-integral})-(\ref{eq:g-norm-integral}) to QMC wave
functions.  Application to GFMC wave functions of the methods used
here will be straightforward and mainly involve additional bookkeeping
similar to that used in Ref.~\cite{brida11} for direct overlap
calculations.  Integrals of the kind considered here are likely to
find their most extensive use in \textit{ab initio} calculations of
coupled-channel scattering and reactions, and a major goal of the work
presented here is to prepare the way for such calculations.

\acknowledgments

I acknowledge useful discussions with C.~A.~Bertulani, I.~Brida,
B.~A.~Brown, C.~R.~Brune, H.~Esbensen, A.~M.~Mukhamedzhanov,
S.~C.~Pieper, and J.~P.~Schiffer.  I thank R.~B.~Wiringa for providing
variational wave functions and much guidance in their use.  This work
was supported by the U.S.  Department of Energy, Office of Nuclear
Physics, under contract No. DE-AC02-06CH11357.  Calculations were
performed on the Fusion computing cluster operated by the Laboratory
Computing Resource Center at Argonne.


\end{document}